\documentclass[letterpaper,english,aps,prb,groupedaddress,twocolumn,showpacs,showkeys]{revtex4-1}
\usepackage[T1]{fontenc}
\usepackage[latin9]{inputenc}
\usepackage{amsmath}
\usepackage{graphicx}
\usepackage{amssymb}

\makeatletter

\providecommand{\LyX}{L\kern-.1667em\lower.25em\hbox{Y}\kern-.125emX\@}
\providecommand{\tabularnewline}{\\}

\@ifundefined{textcolor}{}
{%
 \definecolor{BLACK}{gray}{0}
 \definecolor{WHITE}{gray}{1}
 \definecolor{RED}{rgb}{1,0,0}
 \definecolor{GREEN}{rgb}{0,1,0}
 \definecolor{BLUE}{rgb}{0,0,1}
 \definecolor{CYAN}{cmyk}{1,0,0,0}
 \definecolor{MAGENTA}{cmyk}{0,1,0,0}
 \definecolor{YELLOW}{cmyk}{0,0,1,0}
 }

\usepackage{braket}

\makeatother

\usepackage{babel}

\begin{document}

\title{Dipole trap model for the metallic state in gated silicon-inversion
layers}

\author{T.~H\"ormann}

\affiliation{Institute for Semiconductor Physics, Johannes Kepler University,
4040 Linz, Austria; }

\affiliation{Christian Doppler Labor for Surface Optics, Johannes Kepler University,
4040 Linz, Austria}

\author{G.~Brunthaler}

\email{gerhard.brunthaler@jku.at}

\affiliation{Institute for Semiconductor Physics, Johannes Kepler University,
4040 Linz, Austria}
\begin{abstract}
In order to investigate the metallic state in high-mobility Si-MOS
structures, we have further developed and precised the dipole trap
model which was originally proposed by B.\,L.\ Altshuler and D.\,L.\ Maslov
{[}Phys.\ Rev.\ Lett.\ 82, 145 (1999){]}. Our additional numerical
treatment enables us to drop several approximations and to introduce
a limited spatial depth of the trap states inside the oxide as well
as to include a distribution of trap energies. It turns out that a
pronounced metallic state can be caused by such trap states at appropriate
energies whose behavior is in good agreement with experimental observations.

\end{abstract}

\pacs{71.30.+h; 73.40.Qv; 72.10.Fk}

\keywords{Metal-insulator transition; Si-MOS structures; Scattering model}

\maketitle

\section{introduction}

The discovery of the metal-insulator transition (MIT) in two-dimensional
(2D) electron systems in 1994\citealp{Krav94,Kravchenko1995} has
attracted large attention, as it was in apparent contradiction to
the scaling theory of localization\citealp{Abra79,Abrahams2001} which
states that in the limit of zero temperature, a metallic state should
exist only in three dimensional systems, whereas in two dimensions
disorder should always be strong enough to induce an insulating state.
The MIT in high-mobility n-type silicon inversion layers shows a strong
decrease of resistivity $\rho$ towards low temperature $T$ for high
electron densities, manifesting the metallic region, whereas a strong
exponential increase of $\rho$ demonstrated the insulting regime
for low densities. A similar but weaker behavior was observed in many
other semiconductor systems at low densities and low temperatures
(e.\,g.\ $p$-GaAs,\citealp{Simmons1998} $n$-GaAs,\citealp{Lilly2003}
SiGe,\citealp{Senz2002} AlAs\citealp{Papadakis1998})

Several models were suggested in order to explain the unexpected finding
of metallic behavior in 2D. The most important ones are i) temperature-dependent
screening,\citealp{SternPRL80,GoldPRB86,DasSarma86,Gold2003,DasSarma2005,DasSarma2005b}
ii) quantum corrections in the diffusive regime,\citealp{Finkel84,Castellani84,Punnoose01,Punnoose2005}
and iii) quantum corrections in the ballistic regime.\citealp{Zala01a,Gornyi04}
Numerous argumentations for the different models are given in literature,\citealp{Krav04,Shashkin2005,Dolgopolov2007,Evers2008,Clarke2008}
but a clear decision for one of them could not been drawn yet.

As an alternative, Altshuler and Maslov (AM) introduced the dipol
scattering scenario for Si-MOS structures in which charged trap states
in the oxide layer form dipols together with the image charge of the
screening 2D electrons.\citealp{AM_PhysRevLett.82.145} The interplay
between the gate voltage dependent energetic position of the trap
states and the height of the chemical potential may lead as well to
a metal-insulator transition in that system. It should not be assumed
that the dipol scattering effect is active alone, as the temperature
dependence of screening and quantum corrections will surely contribute
at low temperatures, but the charging of trap states might be the
generator of the particularly large effect in Si-MOS structures. It
is known that the misfit at the silicon/silicon-oxide interface produces
charged defect states in the thermally grown oxide layer.\citealp{Sze_1981_Physics_of_semiconductor_devices,AFS_RevModPhys.54.437,Hori97}
Arguments on the importance of trap states in Si-MOS structures were
also given by Klapwijk and Das Sarma.\citealp{Klapwijk1999}

AM could show within their analytical calculations that a trap level
at energy~$E_{T}$ which is either filled or empty, depending on
its position relative to the Fermi energy~$E_{F}$, can lead to a
critical behavior in electron scattering if $E_{T}$ and $E_{F}$
are degenerate. This dipole trap model is able to explain the main
properties of the metal-insulator transition in gated Si-MOS structures.

For the analytical calculations AM made a number of assumptions. These
are:

a1) the trap states possess a $\delta$-like distribution in energy
(i.\,e.\ have all the same energy),

a2) the spatial density distribution in the oxide is homogeneous,

a3) the states occupied with electrons behave neutral and cause no
scattering of 2D electrons whereas the unoccupied states are positively
charged and lead to scattering (AM work in the hole trap picture,
but we describe occupation in terms of electrons),

a4) a charged trap state is screened by the 2D electrons so that the
resulting electrostatic potential can be described by the trap charge
and an apparent mirror charge with opposite sign on the other side
of the interface,

a5) the scattering efficiency of the 2D electrons is described by
a dipole field of the trap charge and its mirror charge,

a6) a parabolic saddle-point approximation for the effective potential
between the Si/SiO2 interface and the metallic gate was used in order
to perform analytical calculations,

a7) the energy of the trap state $E_{T}$ is fixed relative to the
quantization energy $E_{0}$ of the 2D ground state inside the inversion
potential, and

a8) the chemical potential $\mu$ in the 2D layer has (A) either the
same temperature dependence as in the bulk substrate or (B) as in
a 2D electron system with constant electron density.

In this work, we precise and develop the AM trap model further in
order to better understand the influence of charged traps on the metallic
state in Si-MOS structures. We present detailed numerical calculations
of the temperature and density dependent resistivity $\rho$ due to
electronic scattering in the dipole trap model. Due to the numerical
treatment we were able to drop the approximations a6), a7), and a8)
of the analytic AM model. In addition, we have further extended our
calculations for the more realistic case with energetic broadening
and spatial distribution profile of the defect states, i.\,e.\ dropping
also approximations a1) and a2).

As a result of our calculations, we find good agreement with the calculations
of AM. There are mainly deviations in the overall behavior of the
resistivity at low electron densities and at high temperatures. In
order to understand the approximations of AM, we have also recalculated
the analytical model and were able to formulate it in a simplified
way. Several mathematical terms are rearranged so that the scattering
efficiency is expressed in the same form as in the usual Drude formulation.
In addition, their result contains two integrals, which we could replace
by Fermi-Dirac integrals. Thus the known approximations for the Fermi-Dirac
integrals lead to simple equations for the resistivity~$\rho$ at
low temperatures.

The paper is organized as follows. In Sec.~\ref{sec:Trapmodel} the
analytical formulation of the dipole trap model is given in detail
and we show that within the saddle-point approximation the result
can be written in terms of Fermi-Dirac integrals. Section~\ref{sec:Chemical-potential}
treats the chemical potential and Sec.~\ref{sec:Low-Temp} the analytic
approximations for low temperatures. From the numerical integration,
first results are given in Sec.~\ref{sec:Comparison-of-analytic}
whereas in Sec.~\ref{sec:conduction-band-as-ref} the calculations
are extended for the case that the conduction band is the reference
energy for the trap energy and not the electronic ground state $E_{0}$.
In Sec.~\ref{sec:Spatial-Trap-Profile} the model and the calculation
are extended for a spatial density distribution of the trap states
and in Sec.~\ref{sec:Broadening-Energy} energetic broadening of
trap states is taken into account. Conclusions are drawn in Sec.~\ref{sec:Conclusions}.
In two appendices the behavior of the chemical potential and the ground
state energy of the inversion layer are described in detail. Please
note that we use SI units throughout this work.

\section{Trapmodel\label{sec:Trapmodel}}

\begin{figure}
\includegraphics[width=1\columnwidth]{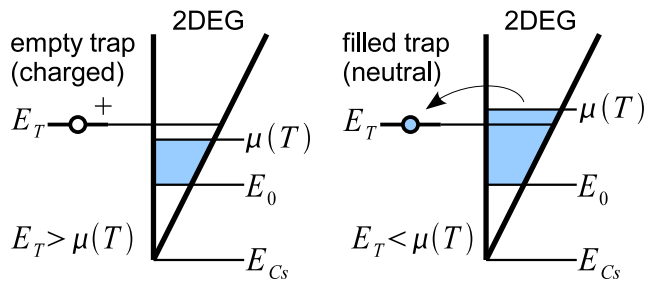}

\caption{\label{fig:OS_trap_charged_or_neutral}Schematic representation of
the trap states together with the 2D electron system in the Si inversion
layer. For $E_{T}>\mu\left(T\right)$ the trap state is positively
charged and scatters electrons in the 2D layer whereas for $E_{T}<\mu\left(T\right)$
the trap is neutral.}

\end{figure}

In the AM model it is assumed that a large number of hole trap states
exists in the oxide at a certain trap energy $E_{T}$. If the trap
energy lies above the chemical potential $\mu$, the trap is empty
(in the electron picture, or has captured a hole in the equivalent
description) and is positively charged, whereas if $E_{T}$ lies below
$\mu$ it is filled with an electron and thus is neutral, see Fig.~\ref{fig:OS_trap_charged_or_neutral}.
Please note that we use in this work the terminology $\mu(T)$ for
the chemical potential, the Fermi energy $E_{F}$ denotes $\mu(T=0)$.

A potential gradient due to an applied gate voltage $V_{g}$ causes
a decrease of the trap energy $E_{T}=E_{Ts}-eV_{\mathrm{ins}}Z/D$,
with the unscreened trap energy $E_{Ts}$ at the oxide semiconductor
(OS) interface, the voltage drop across the oxide (insulator) $V_{\mathrm{ins}}$,
the distance from the OS interface $Z$, and the thickness of the
oxide $D$. In their corresponding equation AM use the symbol $V_{g}$
instead of $V_{\mathrm{ins}}$. . But the total gate voltage $V_{g}$
is equal to the voltage drop across the oxide $V_{\mathrm{ins}}$
plus the voltage drop across the depletion layer $V_{\mathrm{depl}}$.
Later in their paper AM write that the threshold voltage is incorporated
into $V_{g}$, nevertheless they use practically an equation equivalent
to \begin{equation}
V_{\mathrm{ins}}=en_{s}D/\epsilon_{\mathrm{ins}}\epsilon_{0},\label{eq:V-ins-AM}\end{equation}
where $\epsilon_{\mathrm{ins}}$ is the dielectric constant of the
oxide and $\epsilon_{0}$ is the electric field constant. But if the
threshold voltage is incorporated into $V_{g}$, the latter cannot
be used to calculate the slope of the electrical potential within
the oxide, as a part of $V_{g}$ falls off between the OS interface
and the bulk layer. As the charges within the depletion layer (2D
charge density $-en_{\mathrm{depl}}$) also contribute to the gradient
of the potential, we use the equation \begin{equation}
V_{\mathrm{ins}}=e\left(n_{s}+n_{\mathrm{depl}}\right)D/\epsilon_{\mathrm{ins}}\epsilon_{0}\label{eq:V-ins-we}\end{equation}
 together with \begin{equation}
V_{g}=V_{\mathrm{depl}}+V_{\mathrm{ins}}.\end{equation}

According to AM another term has to be added to the trap energy $E_{T}$
which describes the interaction with the two dimensional electron
gas (2DEG) because of the image force. In Ref.~\onlinecite{AM_PhysRevLett.82.145}
this term is given as $-e^{2}/8\pi\epsilon_{\mathrm{ins}}\epsilon_{0}Z$
(translated to the SI unit system). We think there should be a factor
$16$ in the denominator, see for instance Ref.~\onlinecite{Sze_1981_Physics_of_semiconductor_devices}
or Ref.~\onlinecite{AFS_RevModPhys.54.437}. This factor does not
change the results of the trap model qualitatively therefore we write
$-e^{2}/C\pi\epsilon_{\mathrm{ins}}\epsilon_{0}Z$.

\begin{figure}
\includegraphics[width=1\columnwidth]{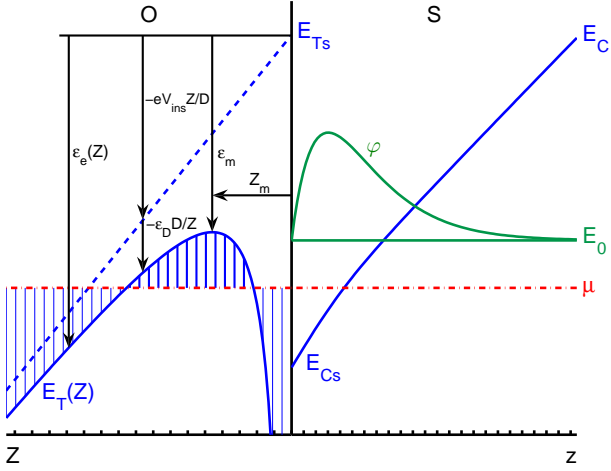}

\caption{\label{fig:OS-overview-1}Oxide semiconductor interface. For simplicity
we use different coordinate systems for the oxide ($Z$) and the semiconductor
side ($z$) so that both, $Z$ and $z$, are positive on their sides.
Trap energy $E_{T}$ which reaches its maximum at $Z=Z_{m}$, unscreened
trap energy $E_{Ts}$ at the OS interface, electrostatic energy $\varepsilon_{e}$,
its components $-eV_{\mathrm{ins}}Z/D$ and $-\varepsilon_{D}D/Z$,
and its maximum $\varepsilon_{m}$, chemical potential $\mu$, ground
state energy of the inversion layer $E_{0}$ and the corresponding
wave function $\varphi$, conduction band edge $E_{C}$ and its value
at the interface $E_{Cs}$. }

\end{figure}

AM define an energy\begin{equation}
\varepsilon_{D}=\frac{e^{2}}{C\pi\epsilon_{\mathrm{ins}}\epsilon_{0}D}\end{equation}
with $C=8$ (AM) or $C=16$ (our assumption). (We use in this work
the notation of capital $E$ for absolute energies and Greek $\varepsilon$
for energy differences.) Finally the trap energy can be written as\begin{gather}
E_{T}\left(Z\right)=E_{Ts}+\varepsilon_{e}\left(Z\right),\\
\varepsilon_{e}\left(Z\right)=-eV_{\mathrm{ins}}\frac{Z}{D}-\varepsilon_{D}\frac{D}{Z},\end{gather}
where the subscript $e$ in $\varepsilon_{e}$ stands for 'electrostatic'.
The shape of $E_{T}\left(Z\right)$ is shown in Fig.~\ref{fig:OS-overview-1}
together with the relevant energy notation.

Without the presence of a magnetic field the probability of a trap
to be charged is given by\begin{equation}
p_{+}\left(Z\right)=\frac{1}{\frac{1}{2}\exp\left(-\frac{E_{T}\left(Z\right)-\mu}{k_{B}T}\right)+1}.\label{eq:p-plus-1}\end{equation}
This formula is similar to the Fermi distribution function, differences
are the minus sign in the exponent and the prefactor $1/2$. A trap
is charged when it is not occupied by an electron, thus the minus
sign. If the trap is charged there are two possibilities for the spin
orientation but only one if it is not charged, leading to the factor
$1/2$. As mentioned before $p_{+}$ is determined by $E_{T}-\mu$
(the vertical lines in Fig.~\ref{fig:OS-overview-1}) and the temperature.

AM assume that a positive charged trap is screened by the electrons
in the 2DEG and the trap forms together with that image charge a dipole.
For the transport scattering cross section $\sigma_{t}$ of this dipole
they found classically\begin{align}
\sigma_{t}\left(\varepsilon,Z\right) & =2.74\left(\frac{e^{2}Z^{2}}{8\pi\epsilon^{*}\epsilon_{0}\varepsilon}\right)^{1/3}\equiv c_{\sigma}\varepsilon^{-1/3}Z^{2/3},\label{eq:sigma-t-1}\\
\rightarrow\quad c_{\sigma} & =2.74\left(\frac{e^{2}}{8\pi\epsilon^{*}\epsilon_{0}}\right)^{1/3},\end{align}
where $Z$ is the distance between the trap and the oxide semiconductor
interface, $\varepsilon$ is the energy of the scattered electron
 relative to the ground state energy of the inversion layer $E_{0}$:
\begin{equation}
\varepsilon=E-E_{0},\end{equation}
and $\epsilon^{*}$ is an effective dielectric constant\begin{equation}
\epsilon^{*}=\frac{\epsilon_{\mathrm{ins}}+\epsilon_{\mathrm{sc}}}{2},\end{equation}
where $\epsilon_{\mathrm{sc}}$ is the dielectric constant of the
semiconductor. We have recalculated Eq.~(\ref{eq:sigma-t-1}) and
got the same result as AM.

The Drude formula together with the Boltzmann equation in relaxation
time approximation yields the resistivity $d\rho\left(Z\right)$ caused
by the charged traps within the layer$\left[Z,Z+dZ\right]$. By integrating
this contributions over the whole oxide $\left[0,D\right]$ one gets\begin{gather}
\rho=\frac{\sqrt{2m_{c}}c_{\sigma}\bar{\varepsilon}^{1/6}\int_{0}^{D}N_{T}^{+}Z^{2/3}dZ}{n_{s}e^{2}},\label{eq:rho-0}\\
\bar{\varepsilon}=\varepsilon_{F0}\left[\int_{0}^{\infty}\frac{1}{4k_{B}T}\left(\frac{\varepsilon}{\varepsilon_{F0}}\right)^{5/6}\cosh^{-2}\left(\frac{\varepsilon-\mu_{E_{0}}}{2k_{B}T}\right)d\varepsilon\right]^{-6},\label{eq:eps_bar_1}\end{gather}
where $m_{c}$ is the conductivity mass of the free electrons within
the inversion layer and $\bar{\varepsilon}$ is an effective electron
energy as used by AM. In the corresponding AM equation the argument
of the $\cosh$ function contains the Fermi energy, but should be
replaced by the chemical potential.%
\footnote{Typo in Eq. 8 of AM paper (D.\,L.~Maslov; private communication)%
} Furthermore $\varepsilon_{F0}=E_{F}-E_{0}$ is the Fermi energy,
$\mu_{E_{0}}=\mu-E_{0}$ is the chemical potential, each relative
to the ground state energy of the inversion layer, and $N_{T}^{+}$
is the 3D density of charged traps. With the 3D trap density $N_{T}\left(Z\right)$
it is given by \begin{equation}
N_{T}^{+}\left(Z\right)=N_{T}\left(Z\right)p_{+}\left(Z\right).\end{equation}

Now we like to present Eq.~(\ref{eq:rho-0}) in a form similar to
the Drude formula\begin{equation}
\rho=\frac{m_{c}}{n_{s}e^{2}\left\langle \tau\right\rangle },\end{equation}
and further find a term for the scattering rate $1/\braket{\tau}$
in the usual form that scattering rate is equal to scattering cross
section times density of scattering centers times velocity of scattered
particles, which leads to one of the basic equations used throughout
this work \begin{equation}
\rho=\frac{m_{c}\sigma_{t}\left(\bar{\varepsilon},\bar{Z}\right)n_{T}^{+}v\left(\bar{\varepsilon}\right)}{n_{s}e^{2}}.\label{eq:rho-1}\end{equation}
Here $\sigma_{t}\left(\bar{\varepsilon},\bar{Z}\right)$ is the transport
scattering cross section from Eq.~(\ref{eq:sigma-t-1}) for the effective
electron energy $\bar{\varepsilon}$ and an effective distance $\bar{Z}$
between the traps and the OS interface, $n_{T}^{+}=\int_{0}^{D}N_{T}^{+}\left(Z\right)dZ$
is the 2D density of charged traps, and $v\left(\bar{\varepsilon}\right)$
is the electron velocity which corresponds with $\bar{\varepsilon}=m_{c}v^{2}/2$.
Comparing Eq.~(\ref{eq:rho-0}) and (\ref{eq:rho-1}) we find\begin{equation}
\bar{Z}^{2/3}=\frac{\int_{0}^{D}N_{T}^{+}\left(Z\right)Z^{2/3}dZ}{\int_{0}^{D}N_{T}^{+}\left(Z\right)dZ}=\frac{\int_{0}^{D}N_{T}^{+}\left(Z\right)Z^{2/3}dZ}{n_{T}^{+}\vphantom{\int_{0}^{D}N_{T}^{+}\left(Z\right)dZ}}\label{eq:Z-bar-1}\end{equation}
for this factor in $\sigma_{t}\left(\bar{\varepsilon},\bar{Z}\right)=c_{\sigma}\bar{\varepsilon}^{-1/3}\bar{Z}^{2/3}$.

In order to calculate the resistivity $\rho$ the knowledge of $n_{T}^{+}$
is not necessary, $n_{T}^{+}$ cancels out with the denominator of
$\bar{Z}^{2/3}$  within $\sigma_{t}\left(\bar{\varepsilon},\bar{Z}\right)$,
AM do not use it. As mentioned in the introduction we are interested
in the metal-insulator transition depending on the electron density
$n_{s}$, i.\,e.\ we want to know the temperature behavior of $\rho$
as a function of $n_{s}$. In this context $n_{T}^{+}$ is very useful
in order to understand on the basis of Eq.~(\ref{eq:rho-1}) that
it contributes the main variations to the resistivity $\rho\left(n_{s},T\right)$
whereas $\sigma_{t}\left(\bar{\varepsilon},\bar{Z}\right)$ and $v\left(\bar{\varepsilon}\right)$
show only weak dependence on $n_{s}$ and $T$. The benefit of Eq.~(\ref{eq:rho-1})
against (\ref{eq:rho-0}) is, that the physical meaning of the terms
is immediately clear.

The integral in the numerator and that in the denominator of $\bar{Z}^{2/3}$
can be treated in quite the same way, so we define\begin{multline}
\Omega_{j}\equiv\int_{0}^{D}N_{T}^{+}\left(Z\right)Z^{j}dZ=\\
=N_{T}\int_{0}^{D}\frac{Z^{j}dZ}{\frac{1}{2}\exp\left(-\frac{E_{T}\left(Z\right)-\mu}{k_{B}T}\right)+1}.\label{eq:Omega-j-1}\end{multline}
In the last step we followed AM and assumed that the trap density
is constant within the oxide, respectively in the region where $p_{+}\left(Z\right)$
does not vanish. Now we can write\begin{gather}
\sigma_{t}\left(\bar{\varepsilon},\bar{Z}\right)=c_{\sigma}\bar{\varepsilon}^{-1/3}\bar{Z}^{2/3}=c_{\sigma}\bar{\varepsilon}^{-1/3}\frac{\Omega_{2/3}}{\Omega_{0}},\\
n_{T}^{+}=\Omega_{0},\\
\rho=\frac{\sqrt{2m_{c}}c_{\sigma}\bar{\varepsilon}^{1/6}\Omega_{2/3}}{n_{s}e^{2}}.\label{eq:rho-2}\end{gather}

\begin{figure}
\includegraphics[width=1\columnwidth]{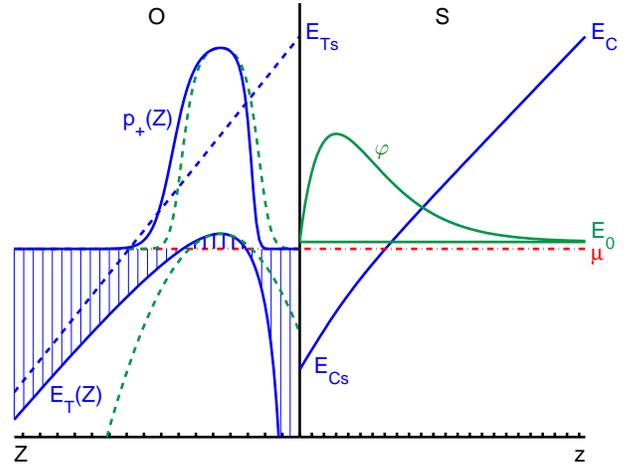}

\caption{\label{fig:SPA}Oxide-semiconductor interface. Trap energy $E_{T}\left(Z\right)$
(full line), its Taylor approximation (dashed line), and the resulting
probabilities $p_{+}\left(Z\right)$ in arbitrary units with $\mu$
as zero point.}

\end{figure}

To be able to calculate the integral which corresponds with $\Omega_{2/3}$
AM expanded the electrostatic energy $\varepsilon_{T}\left(Z\right)$
into a Taylor series about the point $Z_{m}$ where it reaches its
maximum $\varepsilon_{m}$. This procedure is called saddle-point
approximation.\begin{gather}
Z_{m}=D\sqrt{\frac{\varepsilon_{D}}{eV_{\mathrm{ins}}}},\\
\varepsilon_{m}=-2\sqrt{eV_{\mathrm{ins}}\varepsilon_{D}},\\
\varepsilon_{e}\left(Z\right)\simeq\varepsilon_{m}-\varepsilon_{D}\frac{D}{Z_{m}^{3}}\left(Z-Z_{m}\right)^{2},\end{gather}
 see Fig.~\ref{fig:OS-overview-1} and \ref{fig:SPA}. Now (\ref{eq:Omega-j-1})
can be written as\begin{gather}
\Omega_{j}\simeq N_{T}\int_{0}^{D}\frac{Z^{j}dZ}{\frac{1}{2}\exp\left(\frac{\mu_{E_{0}}-\varepsilon_{Ts0}-\varepsilon_{m}+\varepsilon_{D}\frac{D}{Z_{m}^{3}}\left(Z-Z_{m}\right)^{2}}{k_{B}T}\right)+1},\label{eq:Omega-j-2}\\
\varepsilon_{Ts0}=E_{Ts}-E_{0}=\mbox{const.}\end{gather}
AM assume that the energy $E_{Ts}$ relative to the ground state energy
$E_{0}$ is constant, but we believe that rather the conduction band
edge at the interface $E_{Cs}$ has to be used as reference energy,
i.\,e.\ $\varepsilon_{TsCs}=E_{Ts}-E_{Cs}=\mbox{const.}$ This issue
will be further treated in section \ref{sec:conduction-band-as-ref}.

The integrand is a peak around $Z_{m}$ which drops off exponentially
on both sides. In order to come to the same result as AM, we further
apply the following approximations: (i) Because of the exponential
decrease one can integrate from $-\infty$ to $\infty$. (ii) The
integrand is dominated by the denominator, so $Z^{j}\simeq Z_{m}^{j}$
can be set in the numerator.  Now the integrand is symmetric around
$Z_{m}$ and with help of the substitution $\mathcal{Z}=\frac{\varepsilon_{D}D}{Z_{m}^{3}k_{B}T}\left(Z-Z_{m}\right)^{2}$
one gets\begin{multline}
\Omega_{j}=N_{T}Z_{m}^{j+3/2}\sqrt{\frac{k_{B}T}{\varepsilon_{D}D}}\\
\times\int_{0}^{\infty}\frac{\mathcal{Z}^{-1/2}d\mathcal{Z}}{\exp\left(\frac{\mu_{E_{0}}-\varepsilon_{Ts0}-\varepsilon_{m}}{k_{B}T}-\ln2+\mathcal{Z}\right)+1}.\end{multline}

This corresponds to the integral in equation (9c) in Ref.~\onlinecite{AM_PhysRevLett.82.145}
($\mathcal{Z}$ corresponds to $x^{2}$). We brought it into the form
above as it corresponds now to a Fermi-Dirac integral\citealp{Blakemore_Semiconductor_Statistics}\begin{equation}
\mathcal{F}_{k}\left(\eta\right)=\frac{1}{\Gamma\left(k+1\right)}\int_{0}^{\infty}\frac{\mathcal{Z}^{k}d\mathcal{Z}}{\exp\left(\mathcal{Z}-\eta\right)+1},\end{equation}
where $\Gamma$ is the Gamma function. A comparison yields\begin{multline}
\Omega_{j}=N_{T}Z_{m}^{j+3/2}\sqrt{\frac{k_{B}T}{\varepsilon_{D}D}}\\
\times\underbrace{\sqrt{\pi}}_{\Gamma\left(\frac{1}{2}\right)}\mathcal{F}_{-1/2}\left(\ln2+\frac{\varepsilon_{Ts0}+\varepsilon_{m}-\mu_{E_{0}}}{k_{B}T}\right).\label{eq:Omega-j-FDint-1}\end{multline}
On the right hand side $j$ appears only in the exponent of $Z_{m}$,
so within the saddle-point approximation Eq.~(\ref{eq:Z-bar-1})
simplifies to\begin{equation}
\bar{Z}^{2/3}=\frac{\Omega_{2/3}}{\Omega_{0}}\simeq Z_{m}^{2/3}.\end{equation}

\section{Chemical potential\label{sec:Chemical-potential}}

AM described two scenarios for the temperature behavior of the chemical
potential: (A) The chemical potential of the 2DEG and of the Si substrate
coincide. (B) The 2DEG is disconnected form the substrate. For the
case (A) they assumed that the temperature behavior in the 2DEG is
the same as in the bulk. However, they did not take into account that
the chemical potential in the 2DEG is measured against the ground
state energy $E_{0}$ and in the bulk against the conduction or valence
band edge, i.\,e.\ they assumed $E_{0}$ and the band bending to
be fixed.

For (B) AM used an equation analogous to \begin{equation}
\mu_{E_{0}}=k_{B}T\ln\left[\exp\left(\frac{\varepsilon_{F0}}{k_{B}T}\right)-1\right].\label{eq:mu-E0-1}\end{equation}
If only one subband of the inversion layer is occupied (quantum limit),
the Fermi energy relative to its ground state energy is given by\citealp{AFS_RevModPhys.54.437}\begin{equation}
\varepsilon_{F0}=\frac{2\pi\hbar^{2}n_{s}}{g_{s}g_{\mathrm{v2D}}m_{\mathrm{d2D}}},\label{eq:eps-F0}\end{equation}
where $g_{s}$, $g_{\mathrm{v2D}}$, and $m_{\mathrm{d2D}}$ are the
spin degeneracy, the valley degeneracy (for the 2DEG), and the density-of-states
mass (2D) respectively. The two equations above can be derived from\begin{equation}
n_{s}=\underbrace{\frac{g_{s}g_{\mathrm{v2D}}m_{\mathrm{d2D}}}{2\pi\hbar^{2}\vphantom{\exp\left(\frac{E-\mu}{k_{B}T}\right)}}}_{\mathrm{2D\, density\, of\, states}}\int_{E_{0}}^{\infty}\underbrace{\frac{1}{\exp\left(\frac{E-\mu}{k_{B}T}\right)+1}}_{\mathrm{Fermi-Dirac\, distribution}}dE.\end{equation}

\begin{figure}
\includegraphics[width=0.875\columnwidth]{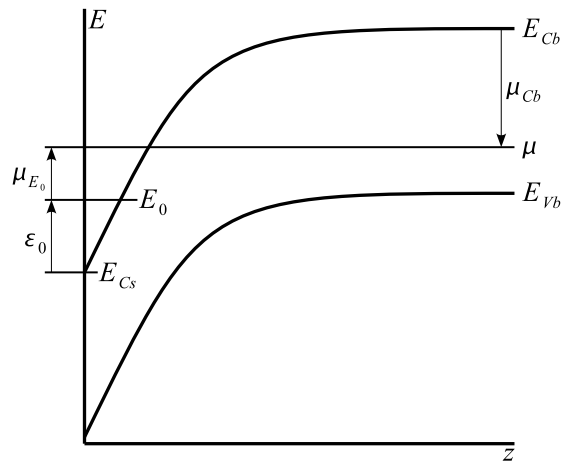}

\caption{\label{fig:OS-overview-2}Band bending and notation on semiconductor
side of the OS interface for thermal equilibrium between 2D layer
and bulk. The electronic ground state energy of the inversion layer
$E_{0}$, the conduction band edge $E_{C}$, its values at the interface
$E_{Cs}$ and in the bulk $E_{Cb}$, the valence band edge $E_{V}$,
and the chemical potential $\mu$ are shown schematically, also the
ground state energy relative to the conduction band edge $\varepsilon_{0}$,
the chemical potential relative to conduction band edge in the bulk
$\mu_{Cb}$ and relative to the ground state energy $\mu_{E_{0}}$
are noted.}

\end{figure}

Our assumptions are shown schematically in Fig.~\ref{fig:OS-overview-2}
(semiconductor side of the OS interface) and are as follows. In thermal
equilibrium there is a single chemical potential $\mu$ throughout
the structure. For a certain temperature $T$, $\mu$ is determined
in the bulk by the (residual) doping density (giving $\mu_{Cb}=\mu-E_{Cb}$,
for details see App.~\ref{sec:Ground-state-energy}). In the inversion
layer the position of $\mu$ relative to $E_{0}$ (i.\,e.\ $\mu_{\varepsilon_{0}}$)
follows directly from the 2D density $n_{s}$.

The band bending $E_{Cb}-E_{Cs}$ adjusts so that\begin{equation}
E_{Cb}-E_{Cs}=\varepsilon_{0}+\mu_{E_{0}}-\mu_{Cb}.\end{equation}
If the band bending is increased, the quantum well gets narrower and
the ground state energy increases. Thus $\varepsilon_{0}$ itself
is a function of the band bending, a self consistent calculation solves
this problem (fix point iteration).

Another possibility is that the 2D electron system is at low temperatures
decoupled from the bulk substrate and no common chemical potential
exists. We do not treat this case in this work, but expect a similar
behavior.

\begin{figure}
\hspace{0.05\columnwidth}(a)\hfill{}\,

\includegraphics[width=0.8\columnwidth]{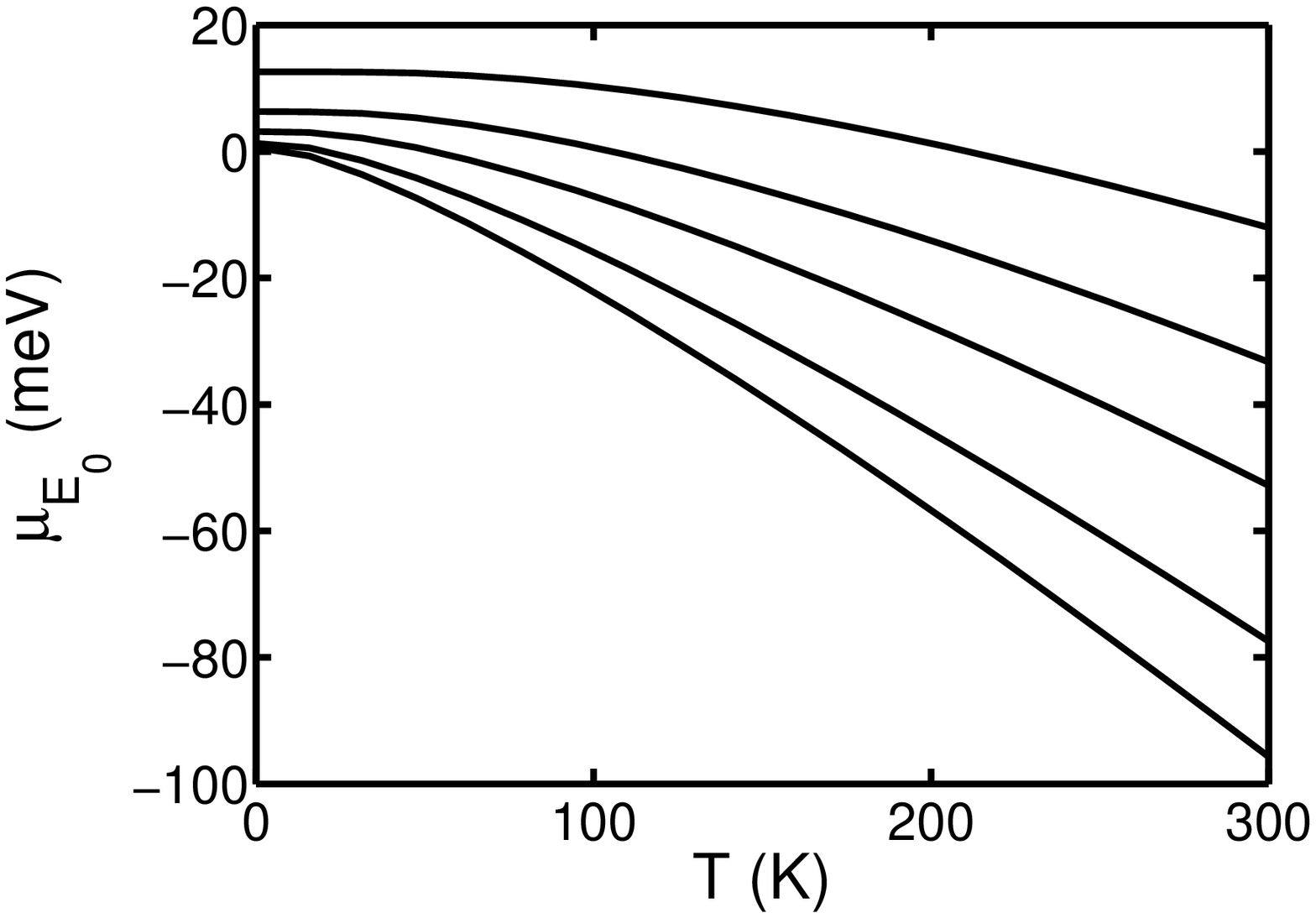}

\hspace{0.05\columnwidth}(b)\hfill{}\,

\includegraphics[width=0.8\columnwidth]{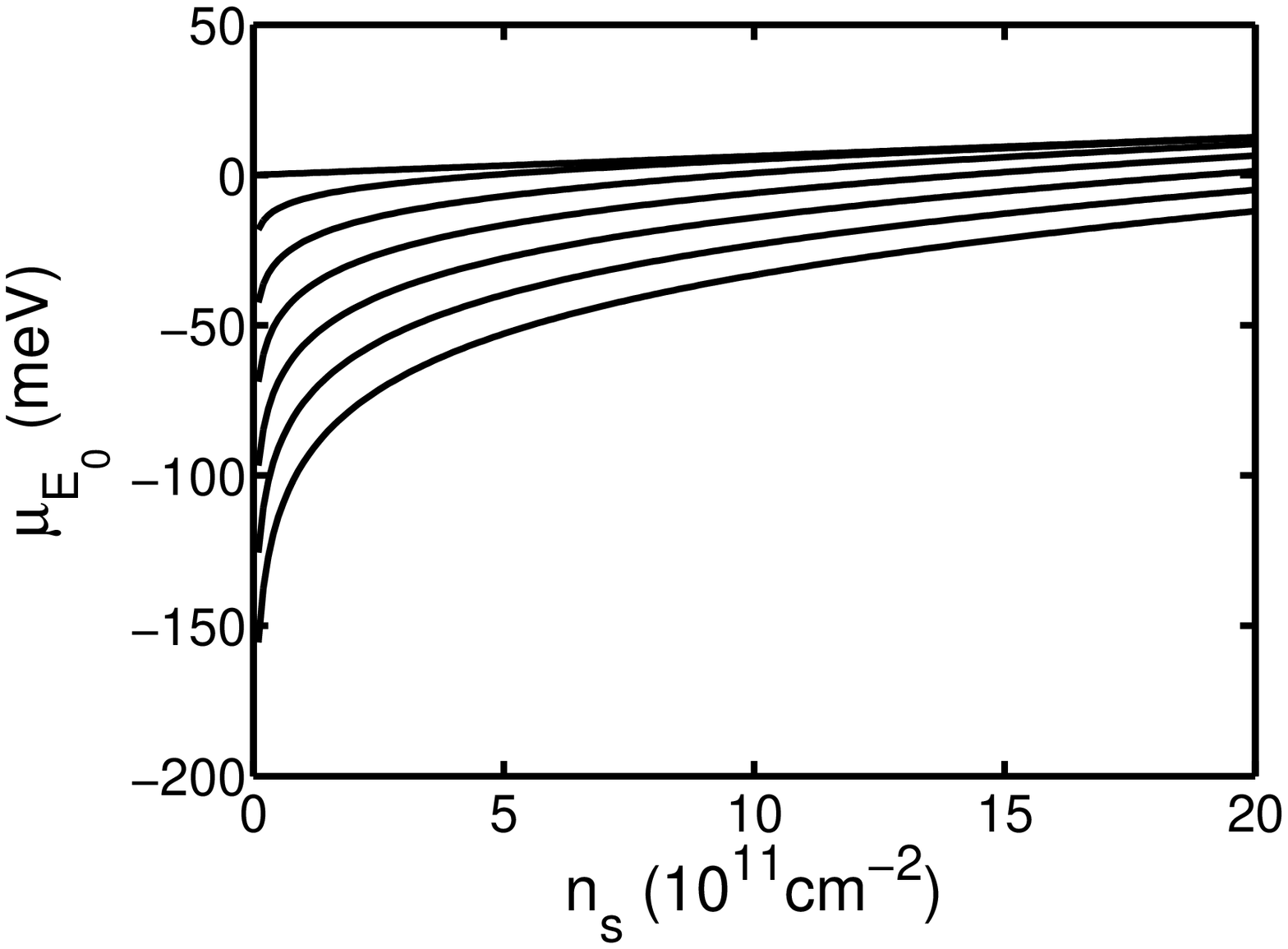}

\caption{\label{fig:Chemical-potential}Chemical potential $\mu$ relative
to the ground state energy $E_{0}$. a) Curves with constant $n_{s}$,
bottom up: $n_{s}=1,2,5,10,20\times10^{11}\,\mathrm{cm}^{-11}$. b)
Curves for constant $T$, top down: $T=0,50,100,150,200,250,300\, K$.}

\end{figure}

Fig.~\ref{fig:Chemical-potential} shows the behavior of $\mu\left(T,n_{s}\right)$
which is crucial for the understanding of the behavior of $\rho\left(T,n_{s}\right)$.
The chemical potential relative to the ground state energy of the
inversion layer $\mu_{E_{0}}$$\left(T,n_{s}\right)$ increases for
decreasing $T$ and for increasing $n_{s}$.

The parameters, which were used here and for the following calculations
are collected in Table \ref{tab:values}. We assumed a \{001\} plane
for the silicon surface.

\begin{table}
\caption{\label{tab:values}Values used for calculations.}

\begin{tabular}{c}
\hline
$g_{s}=2$\tabularnewline
$g_{\mathrm{v2D}}=2$\tabularnewline
$g_{\mathrm{v3D}}=6$\tabularnewline
$m_{e}=9.1094\times10^{-31}\,\mathrm{kg}$\tabularnewline
$m_{\mathrm{d2D}}=0.1905\, m_{e}$\tabularnewline
$m_{z}=0.9163\, m_{e}$\tabularnewline
$m_{\mathrm{de3D}}=0.322\, m_{e}$\tabularnewline
$m_{\mathrm{dh3D}}=0.59\, m_{e}$\tabularnewline
$\epsilon_{\mathrm{sc}}=11.9$\tabularnewline
$\epsilon_{\mathrm{ins}}=3.9$\tabularnewline
$\rightarrow\epsilon^{*}=7.9$\tabularnewline
$D=200\,\mathrm{nm}$\tabularnewline
$C=16$\tabularnewline
$\rightarrow\varepsilon_{D}=0.4615\,\mathrm{meV}$\tabularnewline
$N_{A}=2\times10^{15}\,\mathrm{cm}^{-3}$\tabularnewline
$N_{D}=0$\tabularnewline
$\varepsilon_{\mathrm{g0Si}}=1.17\,\mathrm{eV}$\tabularnewline
$\alpha_{\mathrm{Si}}=4.73\times10^{-4}\,\mathrm{eVK}^{-1}$\tabularnewline
$\beta_{\mathrm{Si}}=636\,\mathrm{K}$\tabularnewline
\hline
\end{tabular}
\end{table}

\section{Approximation for low temperatures\label{sec:Low-Temp}}

For given values $T$ and $n_{s}$ we calculate the resistivity $\rho$
with help of the equations (\ref{eq:rho-2}) and (\ref{eq:Omega-j-FDint-1}).
Beside the explicit temperature dependence also the chemical potential
$\mu_{E_{0}}$ and the effective electron energy $\bar{\varepsilon}$
are functions of $T$.

As a first step we replace the integral in (\ref{eq:eps_bar_1}) by
a Fermi-Dirac integral. For the Fermi distribution function\begin{equation}
f=\frac{1}{\exp\left(\frac{E-\mu}{k_{B}T}\right)+1}=\frac{1}{\exp\left(\frac{\varepsilon-\mu_{E_{0}}}{k_{B}T}\right)+1}\end{equation}
the following identity holds,\begin{equation}
-\frac{\partial f\left(\varepsilon\right)}{\partial\varepsilon}=\frac{1}{4k_{B}T}\cosh^{-2}\left(\frac{\varepsilon-\mu_{E_{0}}}{2k_{B}T}\right).\end{equation}
So Eq.~(\ref{eq:eps_bar_1}) can be written as\begin{equation}
\bar{\varepsilon}=\varepsilon_{F0}\left[\int_{0}^{\infty}\left(\frac{\varepsilon}{\varepsilon_{F0}}\right)^{5/6}\left(-\frac{\partial f\left(\varepsilon\right)}{\partial\varepsilon}\right)d\varepsilon\right]^{-6}.\end{equation}
An integration by parts yields\begin{align}
\bar{\varepsilon} & =\varepsilon_{F0}^{6}\left[\frac{5}{6}\int_{0}^{\infty}\varepsilon^{-1/6}f\left(\varepsilon\right)d\varepsilon\right]^{-6},\nonumber \\
\bar{\varepsilon} & =\varepsilon_{F0}\left(\frac{\varepsilon_{F0}}{k_{B}T}\right)^{5}\nonumber \\
 & \quad\times\left[\frac{5}{6}\int_{0}^{\infty}\frac{\left(\frac{\varepsilon}{k_{B}T}\right)^{-1/6}}{\exp\left(\frac{\varepsilon-\mu_{E_{0}}}{k_{B}T}\right)+1}d\left(\frac{\varepsilon}{k_{B}T}\right)\right]^{-6},\nonumber \\
\bar{\varepsilon} & =\varepsilon_{F0}\left(\frac{\varepsilon_{F0}}{k_{B}T}\right)^{5}\left[\Gamma\left(\frac{11}{6}\right)\mathcal{F}_{-1/6}\left(\frac{\mu_{E_{0}}}{k_{B}T}\right)\right]^{-6}.\label{eq:epsilon-bar-6}\end{align}

For low temperatures we can take advantage of the behavior of $\mu_{E_{0}}\left(T\right)$
and $\bar{\varepsilon}\left(T\right)$ for $T\rightarrow0$. It can
easily be shown that for constant $n_{s}$ and therefore constant
$\varepsilon_{F0}$ all derivatives vanish at this point. \begin{align}
\left.\frac{d^{n}\mu_{E0}\left(T\right)}{dT^{n}}\right|_{T\rightarrow0} & =0\quad\mathrm{for\,}n\ge1,\label{eq:mu_E0_T0_1}\\
\left.\frac{d^{n}\bar{\varepsilon}\left(T\right)}{dT^{n}}\right|_{T\rightarrow0} & =0\quad\mathrm{for\,}n\ge1.\end{align}
So $\mu_{E_{0}}\left(T\right)$ and $\bar{\varepsilon}\left(T\right)$
are very flat functions at $T\rightarrow0$, for $k_{B}T\ll\varepsilon_{F0}$
they can be approximated by\begin{align}
\mu_{E_{0}}\left(T\right) & \simeq\varepsilon_{F0},\\
\bar{\varepsilon}\left(T\right) & \simeq\varepsilon_{F0}.\end{align}
See appendix \ref{sec:behavior-of-the-chemical-potential} for details.

The Fermi-Dirac integral $\mathcal{F}_{j}\left(\eta\right)$ can be
approximated by\citealp{Blakemore_Semiconductor_Statistics}\begin{equation}
\mathcal{F}_{j}\left(\eta\right)\simeq\frac{\eta^{j+1}}{\Gamma\left(j+2\right)}\quad\mathrm{for}\,\eta\gg0,\end{equation}
which is the first term of an asymptotic series derived with help
of a Sommerfeld expansion,\citealp{Ashcroft_Mermin_Solid_State_Phys}
and by\begin{equation}
\mathcal{F}_{j}\left(\eta\right)\simeq\exp\eta\quad\mathrm{for}\,\eta\ll0.\end{equation}
For $T\rightarrow0$ the argument of the Fermi-Dirac integral in Eq.~(\ref{eq:Omega-j-FDint-1})
increases beyond any border. Which approximation for the Fermi-Dirac
integral is applicable depends on the sign of $\varepsilon_{Ts0}+\varepsilon_{m}-\varepsilon_{F0}$.
(Here $\mu_{E_{0}}$ is replaced by $\varepsilon_{F0}$ as by definition
$\mu_{E_{0}}\rightarrow\varepsilon_{F0}$ for $T\rightarrow0$.) This
is conform with AM's definition of the transition point, $\left(\varepsilon_{Ts0}+\varepsilon_{m}-\varepsilon_{F0}\right)/k_{B}T=0$.
Accordingly we define\begin{align*}
\varepsilon_{mF} & =\varepsilon_{Ts0}+\varepsilon_{m}-\varepsilon_{F0},\\
\varepsilon_{mF} & >0\quad\rightarrow\mathrm{insulating,}\\
\varepsilon_{mF} & <0\quad\rightarrow\mathrm{metallic,}\\
\varepsilon_{mF} & =0\quad\rightarrow\mathrm{transition\, point}\end{align*}
and a critical density\begin{equation}
n_{sc}=\left.n_{s}\right|_{\varepsilon_{mF}=0}.\end{equation}

Applying the appropriate approximations results in\begin{equation}
\Omega_{j}\simeq\begin{cases}
2N_{T}Z_{m}^{j+3/2}\sqrt{\frac{k_{B}T\ln2}{\varepsilon_{D}D}+\frac{\varepsilon_{mF}}{\varepsilon_{D}D}} & \mathrm{for}\,\varepsilon_{mF}>0\\
2N_{T}Z_{m}^{j+3/2}\sqrt{\frac{k_{B}T\pi}{\varepsilon_{D}D}\exp\left(\frac{\varepsilon_{mF}}{k_{B}T}\right)} & \mathrm{for}\,\varepsilon_{mF}<0\\
\underbrace{\mathcal{F}_{-1/2}\left(\ln2\right)}_{\simeq0.891}N_{T}Z_{m}^{j+3/2}\sqrt{\frac{k_{B}T\pi}{\varepsilon_{D}D}} & \mathrm{for}\,\varepsilon_{mF}=0.\end{cases}\end{equation}
Please note an interesting behavior. When setting $\varepsilon_{mF}=0$
in the first two equations for $T>0$ they converge neither into each
other nor into the third one. This apparent discrepancy can be understood,
as $\left|\varepsilon_{mF}\right|$ gets smaller and smaller the maximum
temperature where the approximations for the Fermi-Dirac integral
are just applicable also gets smaller and smaller and finally vanishes
for $\varepsilon_{mF}=0$. Indeed for $\varepsilon_{mF}=0$ and $T\rightarrow0$
the three cases yield the same result, i.\,e.\ $\Omega_{j}=0$.

\section{Comparison of analytic and numerical results\label{sec:Comparison-of-analytic}}

\begin{figure*}
\hspace{0.1\textwidth}(a)\hfill{}(b)\hfill{}\hspace{0.1\textwidth}

\includegraphics[width=0.4\textwidth]{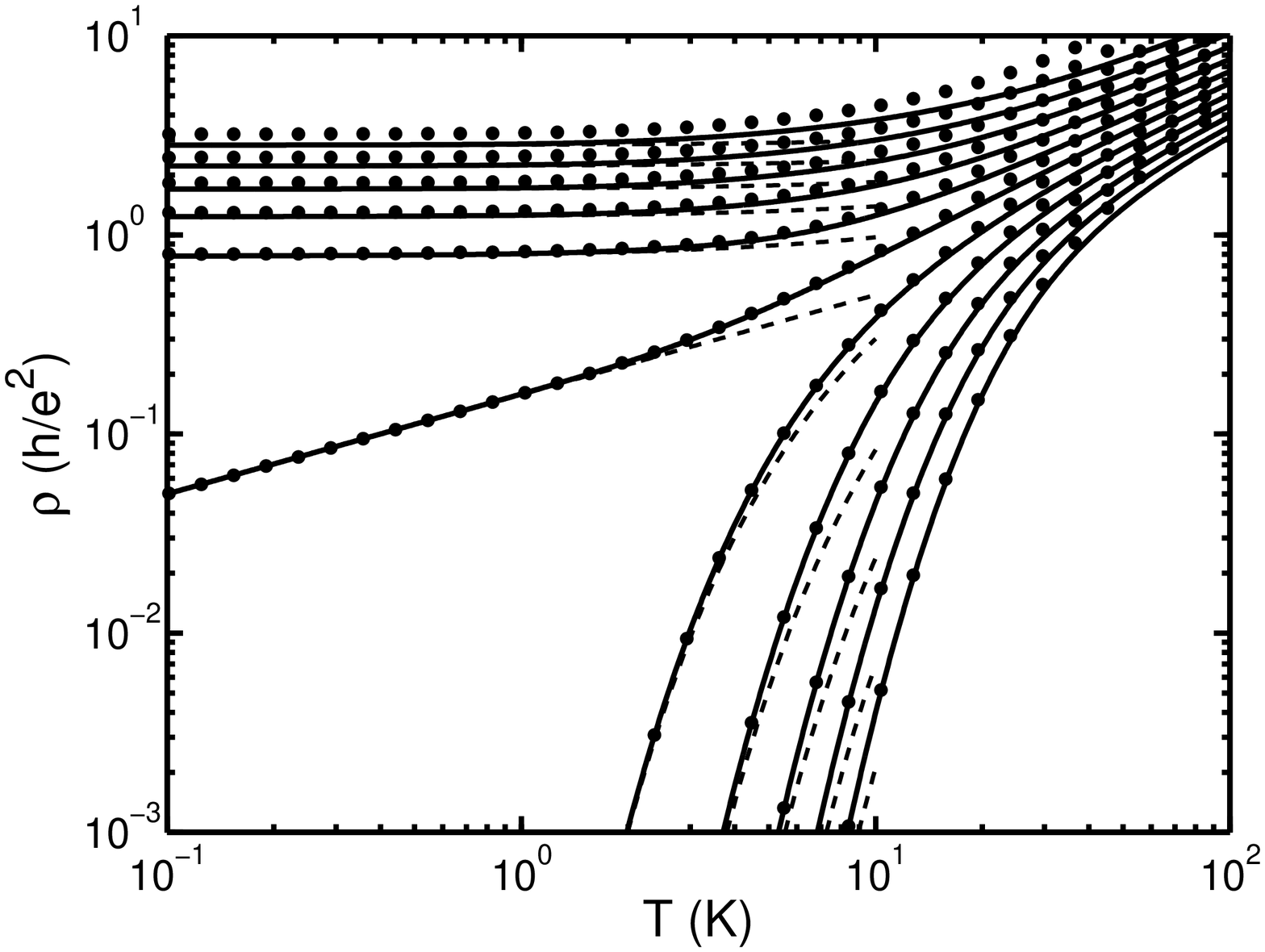}\includegraphics[width=0.4\textwidth]{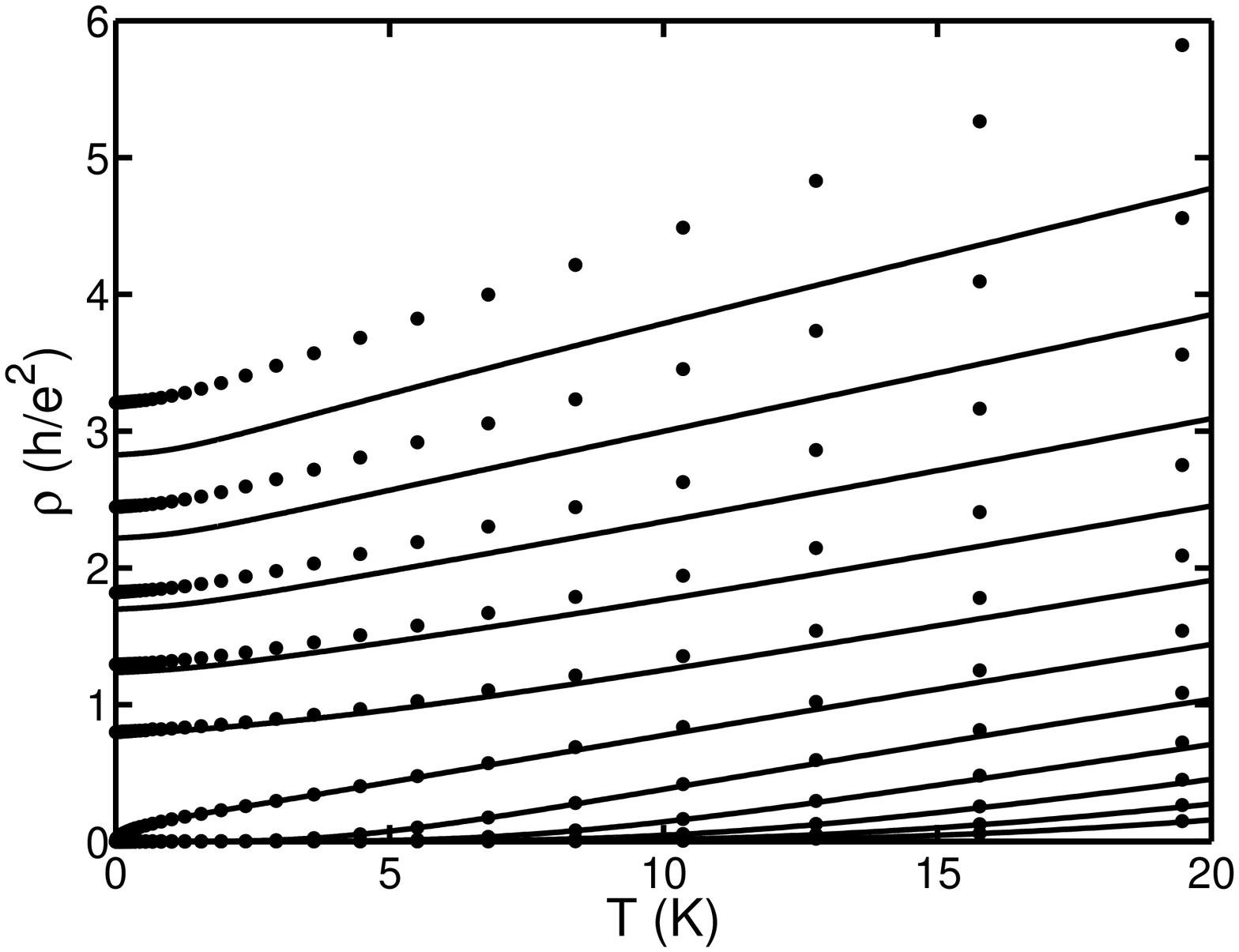}

\hspace{0.1\textwidth}(c)\hfill{}(d)\hfill{}\hspace{0.1\textwidth}

\includegraphics[width=0.4\textwidth]{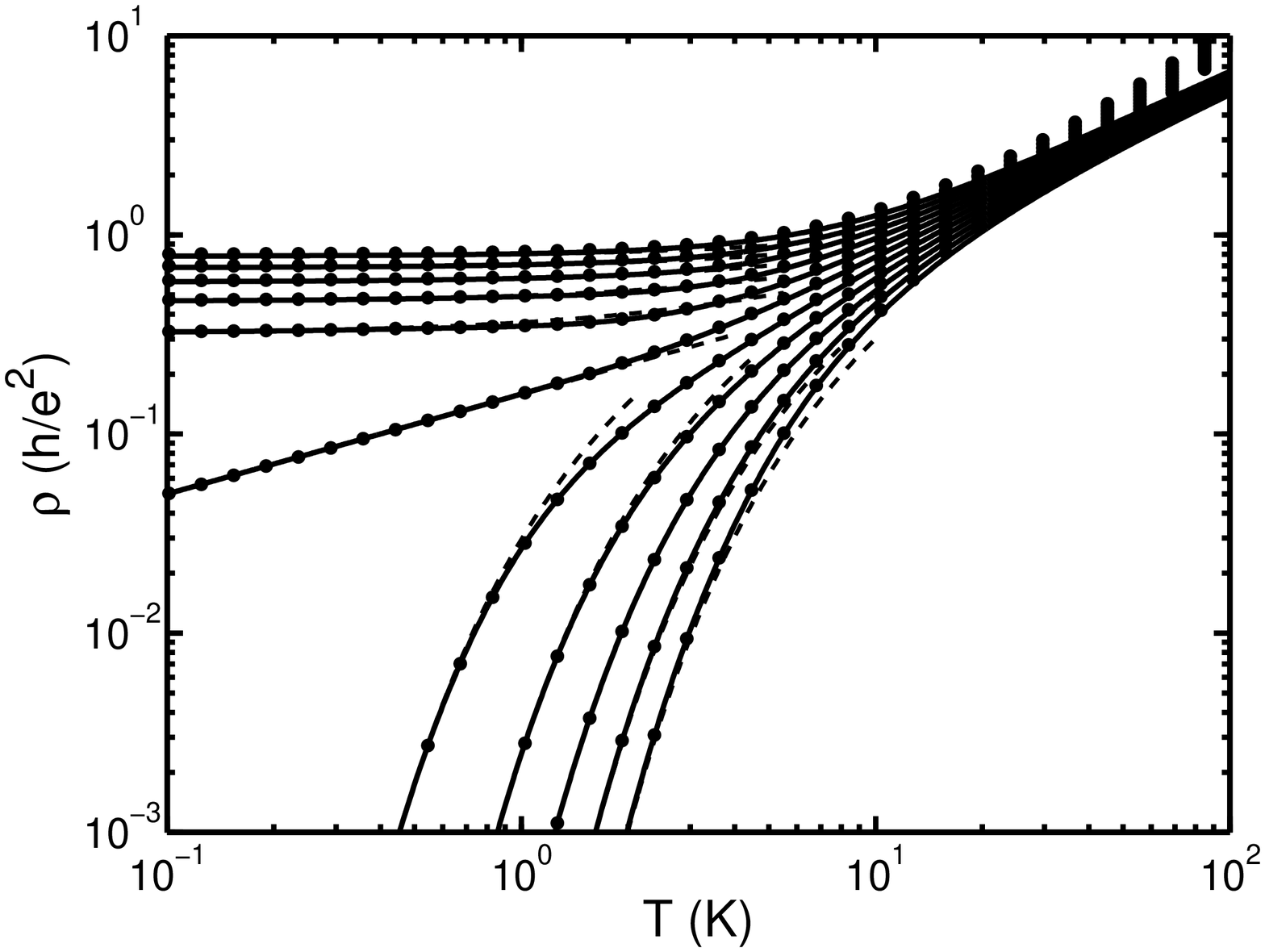}\includegraphics[width=0.4\textwidth]{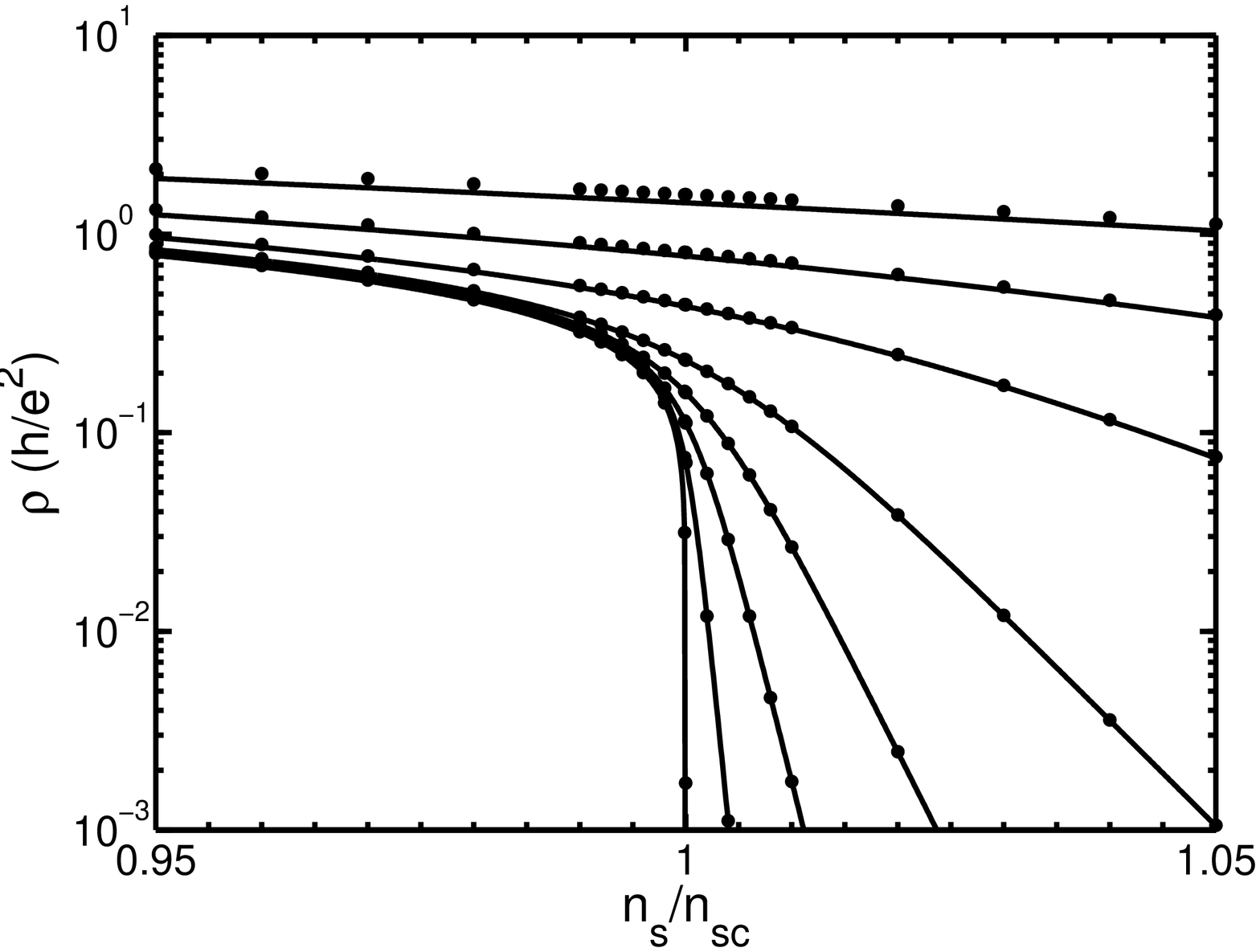}

\caption{\label{fig:SPA_numInt_lowTemp}Behavior of the resistivity $\rho$
depending on the temperature $T$ and the 2D electron density in the
inversion layer $n_{s}$. Full lines: saddle-point approximation,
point symbols: numerical integration, dashed lines: low temperature
approximations. On the curves in a) - c) $n_{s}$ and in d) $T$ is
constant. Critical density $n_{sc}=10^{11}\,\mbox{cm}^{-2}$. a) logarithmic
and b) linear display of $\rho(T)$ with parameters: $n_{s}=0.75,\dots,1.25\times n_{sc}$
in steps of $0.05\times n_{sc}$ in top down order for the individual
curves, c) logarithmic view of $\rho(T)$ with top down parameters:
$n_{s}=0.95,\dots,1.05\times n_{sc}$ in steps of $0.01\times n_{sc}$,
d) $\rho(n_{s})$ with parameters $T=0,\,0.2,\,0.5,\,1,\,2,\,5,\,10,\,20\,\mathrm{K}$
in bottom up order.}

\end{figure*}

In order to get rid of the restrictions which came from the saddle-point
approximation we have performed numerical calculations of the integrals
$\Omega_{j}$ (Eq.~(\ref{eq:Omega-j-2})). In Fig.~\ref{fig:SPA_numInt_lowTemp}
we show the resistivity $\rho$ depending on the temperature $T$
and the 2D electron density in the inversion layer $n_{s}$ calculated
with the help of the saddle-point approximation (full lines), the
approximation for low temperatures (dashed lines), and the numerical
integration (markers). We chose $\varepsilon_{Ts0}\simeq42.02\,\mathrm{meV}$
in order to get $n_{sc}=10^{11}\mathrm{cm^{-11}}$, for the 3D trap
density we assumed $N_{T}=10^{18}\,\mathrm{cm}^{-3}$. The value for
$N_{T}$ seems to be quite high, but only in a very narrow layer the
traps will indeed be charged (where the trap states are above the
chemical potential) and contribute to scattering. A further limitation
of the available trap states into a narrow region besides the interface
will follow later in this work.

The critical behavior of $\rho\left(T\right)$ versus $n_{s}$ for
the homogeneous trap density $N_{T}$ can be seen most clearly on
the double logarithmic Figs.~\ref{fig:SPA_numInt_lowTemp}a) and
c). We added \ref{fig:SPA_numInt_lowTemp}b) as it is in this form
directly comparable with Fig.~1b) in the work of AM.\citealp{AM_PhysRevLett.82.145}

We like to explain the shape of $\rho\left(T,n_{s}\right)$ starting
from a value $n_{s}>n_{sc}$. The maximum of $\varepsilon_{T}\left(Z\right)$
lies below $\mu$, therefore the $p_{+}\left(Z\right)$-peak is very
small and narrow. When $n_{s}$ is decreased $\mu$ drops off (see
Fig.~\ref{fig:Chemical-potential}) and the resistivity $\rho$ rises
fast because $-\left(E_{T}-\mu\right)/k_{B}T$ is the exponent in
the denominator of $p_{+}$. It rises the faster the smaller the temperature
is. When $\mu$ reaches the maximum of the trap energy $E_{T}$ the
$p_{+}$-peak has a height of $2/3$, therefore further decreasing
of $n_{s}$ and $\mu$ cannot increase the height of the peak appreciable
(on a logarithmic scale, which covers some orders of magnitude), only
the width.

For increasing temperature $T$ also the chemical potential $\mu$
measured against $E_{0}$ decreases which results in increasing resistivity
$\rho$. Additionally the $p_{+}\left(Z\right)$-peak is broadened
because the Fermi distribution function declines over several $k_{B}T$.

The saddle-point approximation works best for small temperatures $T$
and densities $n_{s}>n_{sc}$, in this domain the $p_{+}\left(Z\right)$-peak
is very narrow and therefore the Taylor approximation of the trap
energy is accurate within the peak. For large $T$ and/or $n_{s}<n_{sc}$
the deviations of the saddle-point approximation from the numerical
calculations is visible in Fig.~\ref{fig:SPA_numInt_lowTemp}.

It should be noted that according to the calculations, the resistivity
$\rho$ drops to arbitrarily low values in the metallic regime. This
is caused by the fact that electron scattering is taken into account
only by the trap states at a single trap energy. If this trap energy
is below the Fermi energy, with decreasing $T$ the number of charged
scattering centers goes to zero. Only if other scattering effects,
like residual impurities, surface roughness, acceptor states in the
depletion layer, etc.\ are included, the low-$T$ resistivity would
be limited. As will be seen later, an energetic broadening of the
trap states will have a similar effect.

\section{\label{sec:conduction-band-as-ref}Conduction band as reference energy}

If the bands in the semiconductor and in the oxide are bent due to
an applied gate voltage all energies move up and down with the bands.
The energetic position of the trap states $E_{Ts}$ should thus be
fixed relative to the conduction band edge and not to the ground state
energy $E_{0}$ of the inversion layer as assumed by AM.

From Eq.~(\ref{eq:p-plus-1}) we see that $E_{T}-\mu=E_{Ts}+\varepsilon_{e}\left(Z\right)-\mu$
determines the probability $p_{+}$ of a trap to be charged. So if
we measure $E_{Ts}$ against $E_{Cs}$ we also have to know $\mu_{Cs}=\mu-E_{Cs}$.
We find\begin{equation}
\mu_{Cs}=\mu-E_{0}+E_{0}-E_{Cs}=\mu_{E_{0}}+\varepsilon_{0},\end{equation}
where $\varepsilon_{0}=E_{0}-E_{Cs}$ is the ground state energy relative
to the conduction band edge at the OS interface.

The chemical potential $\mu_{E_{0}}$ follows from Eq.~(\ref{eq:mu-E0-1}),
but an accurate calculation of $\varepsilon_{0}$ is rather complex.
For simplicity we follow the calculation method of Ando, Fowler and
Stern (AFS)\citealp{AFS_RevModPhys.54.437} and neglect the exchange
interaction and correlation effects and use the Ritz variational principle.
(In the mentioned article also more sophisticated methods for the
calculation of $\varepsilon_{0}$ are given.)

For convenience we introduce a new coordinate system $z=-Z$, i.\,e.\ the
$z$-axis is perpendicular to the OS interface, positive $z$-values
correspond with the semiconductor side. For the electrons in the inversion
layer the bent conduction band of the semiconductor together with
the step at the interface builds the quantum well. We use the Fang-Howard
envelope wave function according to AFS \citealp{Fang_Howard_1966}\begin{equation}
\varphi\left(z,b\right)=\begin{cases}
\sqrt{\frac{b^{3}}{2}}z\exp\left(-\frac{bz}{2}\right) & \mathrm{for}\, z\ge0\\
0 & \mathrm{for}\, z<0.\end{cases}\end{equation}
The parameter $b$ is varied in order to make the total energy per
electron minimal. For the potential several approximations are taken,
see App.~\ref{sec:Ground-state-energy} for details.

\begin{figure}
\includegraphics[width=0.8\columnwidth]{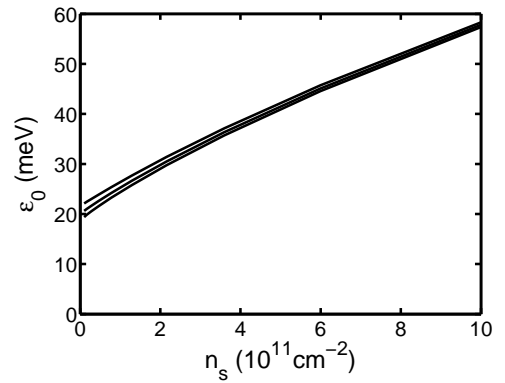}

\caption{\label{fig:Ground-state-energy}Ground state energy of the inversion
layer $\varepsilon_{0}$ versus the density of electrons in the inversion
layer $n_{s}$ for three temperatures, top-down: $T=0,200,300\, K$.}

\end{figure}

\begin{figure*}
\hspace{0.1\textwidth}(a)\hfill{}(b)\hfill{}\hspace{0.1\textwidth}

\includegraphics[width=0.4\textwidth]{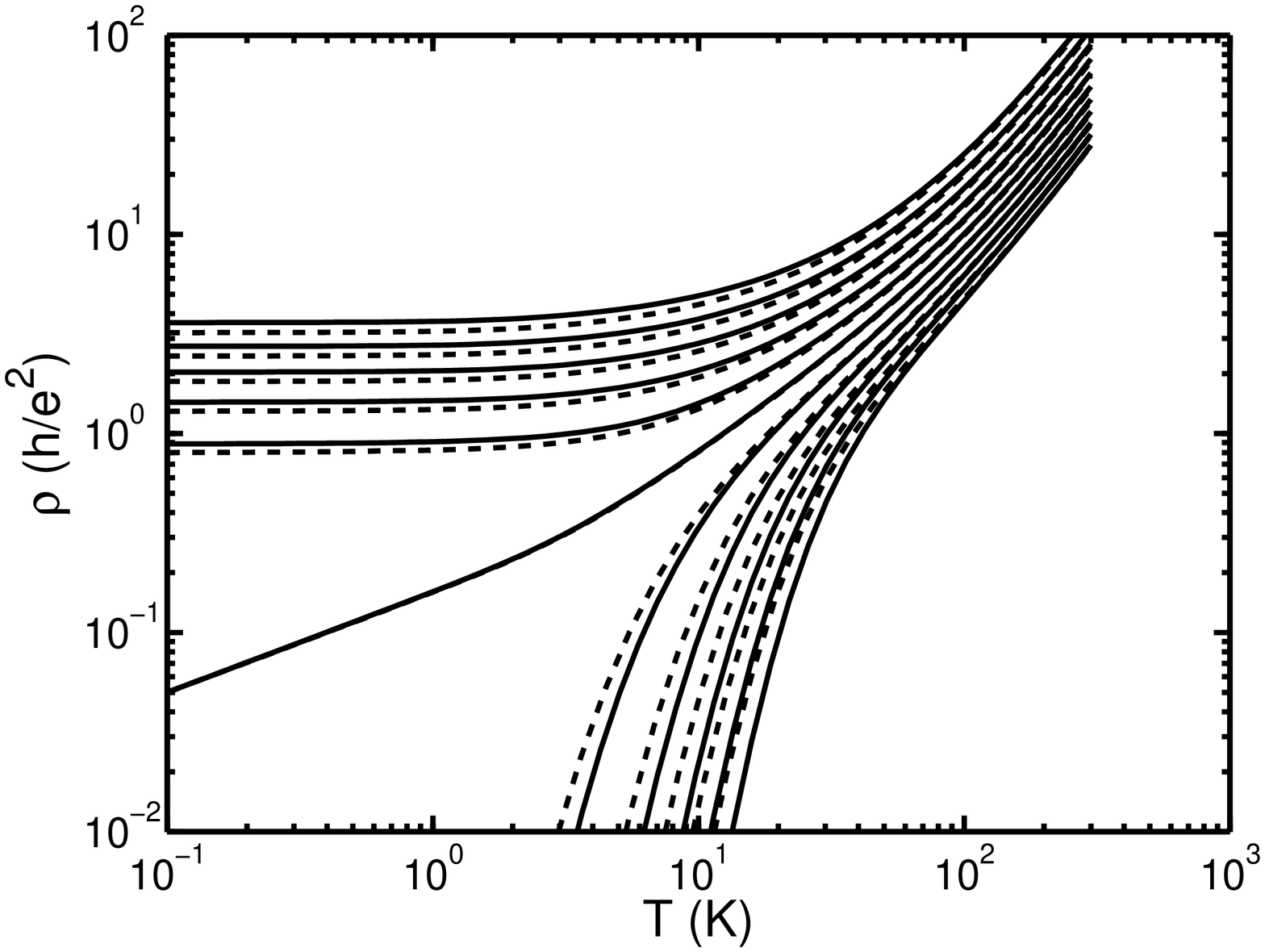}\includegraphics[width=0.4\textwidth]{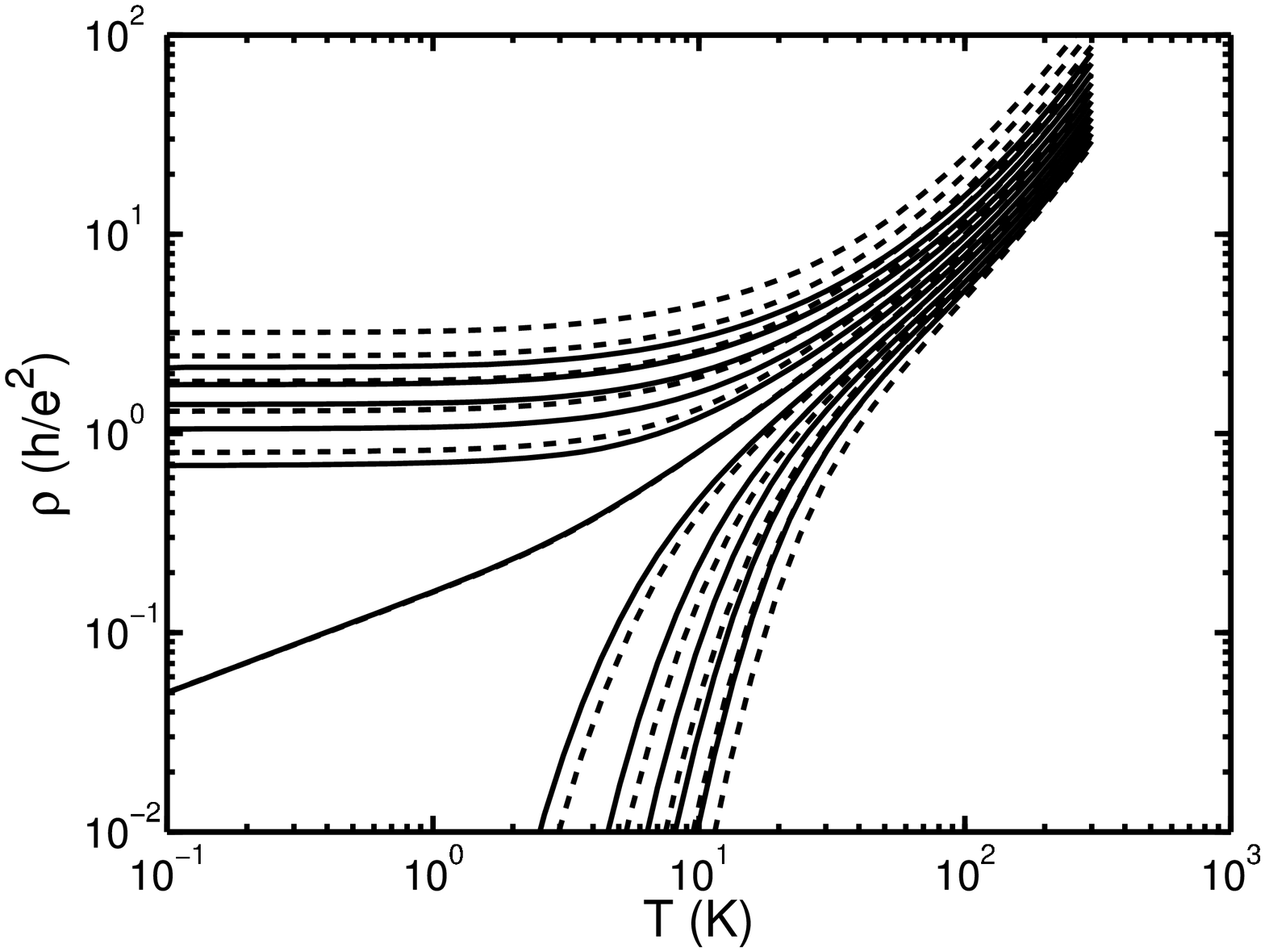}

\hspace{0.1\textwidth}(c)\hfill{}(d)\hfill{}\hspace{0.1\textwidth}

\includegraphics[width=0.4\textwidth]{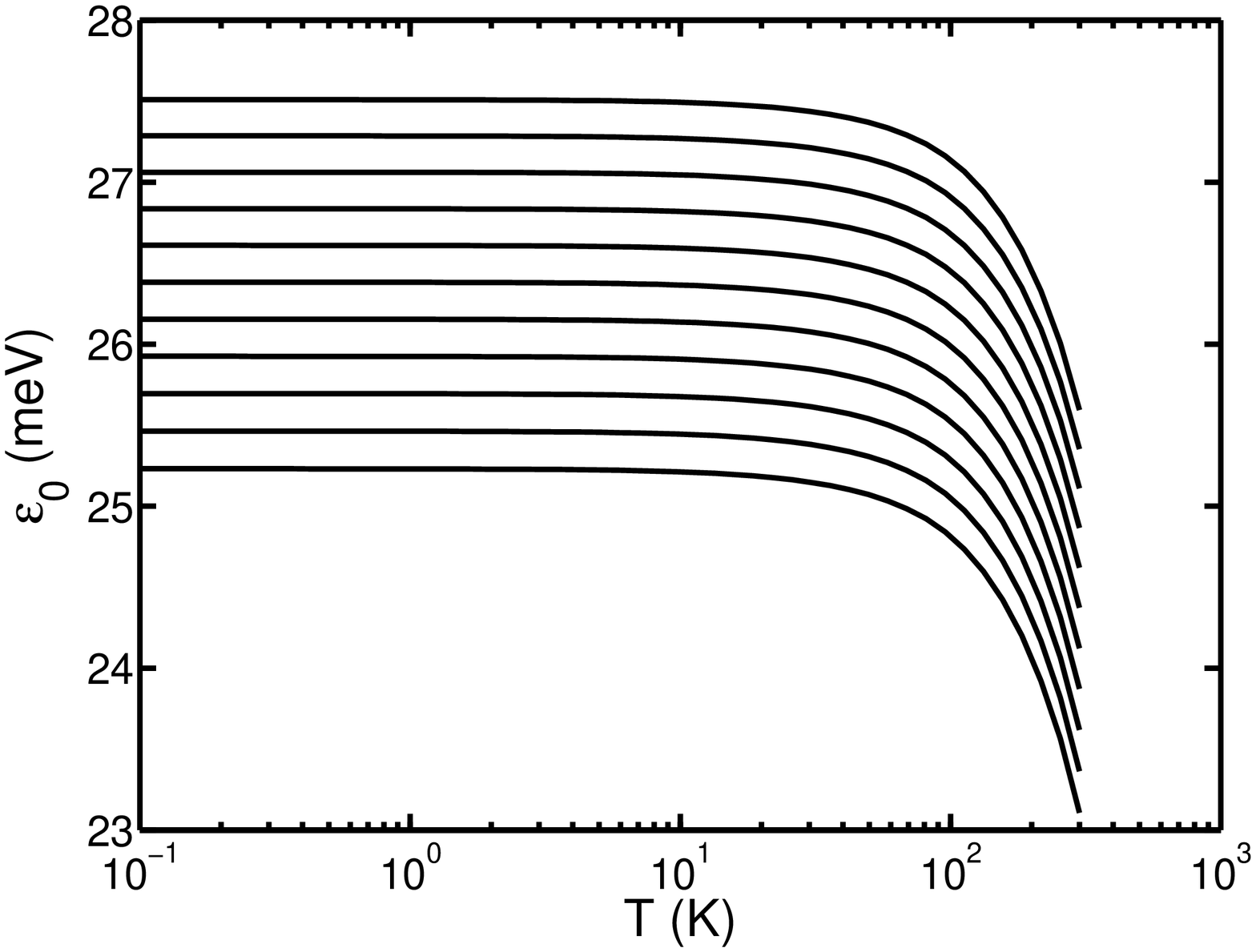}\includegraphics[width=0.4\textwidth]{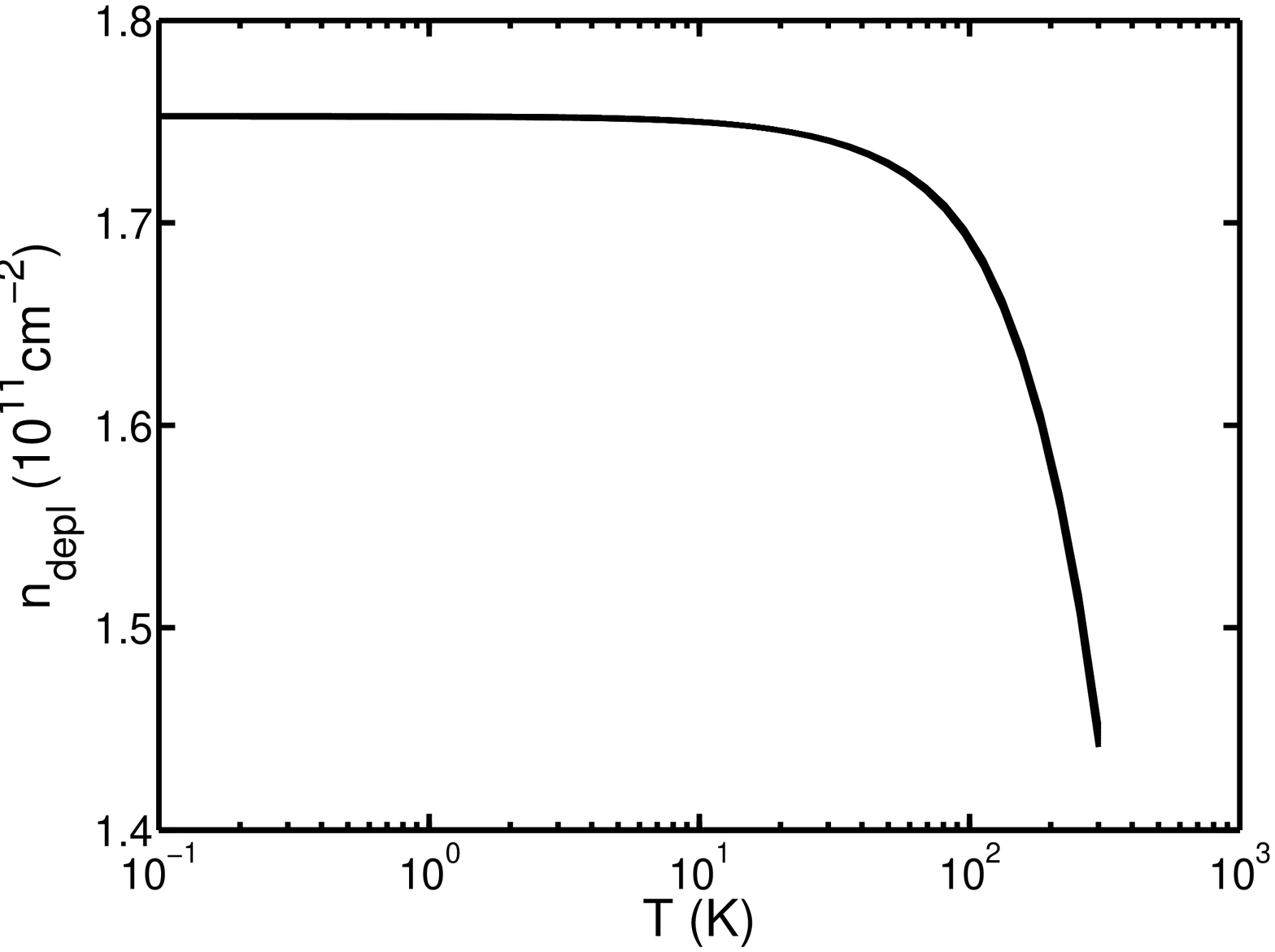}

\caption{\label{fig:Influence-of-E0}Resistivity $\rho\left(T,n_{s}\right)$
for the potential across the oxide layer a) $V_{\mathrm{ins}}\propto n_{s}$
with $N_{T}=10^{18}\,\mathrm{cm}^{-3}$ and $\varepsilon_{TsCs}\simeq68.4\,\mathrm{meV}$
and b) $V_{\mathrm{ins}}\propto n_{s}+n_{\mathrm{depl}}$ with $N_{T}=2.98\times10^{18}\,\mathrm{cm}^{-3}$
and $\varepsilon_{TsCs}\simeq95.7\,\mathrm{meV}$. The parameters
are chosen to give a critical density of $n_{sc}=10^{11}\,\mbox{cm}^{-2}$
for both cases. At the individual curves $n_{s}=0.75,\dots,1.25\times n_{sc}$
in steps of $0.05\times n_{sc}$ from top to bottom, full lines represent
$\varepsilon_{TsCs}=\mathrm{const.}$ (realistic case), dashed lines
$\varepsilon_{Ts0}=\mathrm{const.}$ (for comparison). In c) the ground
state energy of the inversion layer $\varepsilon_{0}\left(T,n_{s}\right)$
is shown for same $n_{s}$ values as before, but now assigned bottom
up. In d) the depletion density $n_{\mathrm{depl}}\left(T,n_{s}\right)$
is shown, but the $n_{s}$ dependence vanishes within the line width.}

\end{figure*}

In Fig.~\ref{fig:Ground-state-energy} the ground state energy $\varepsilon_{0}$
versus the electron density $n_{s}$ is shown, $\varepsilon_{0}$
decreases with decreasing $n_{s}$. Now we hold the difference between
trap energy and conduction band edge at the interface $\varepsilon_{TsCs}=E_{Ts}-E_{Cs}=E_{Ts}-\mu+\mu_{E_{0}}+\varepsilon_{0}$
constant (instead of $\varepsilon_{Ts0}$ as before). When $n_{s}$
is decreased not only $\mu_{E_{0}}$ but also $\varepsilon_{0}$ decreases,
thus $\mu$ drops off faster against $E_{Ts}$ and the transition
is more abrupt. This can be seen in Fig.~\ref{fig:Influence-of-E0}a)
where results for $\varepsilon_{TsCs}=\mathrm{const.}$ (full lines)
and $\varepsilon_{Ts0}=\mathrm{const.}$ (dashed lines) are compared.
In both cases $\Omega_{j}$ was calculated numerically. The critical
curves $n_{s}=n_{sc}$ coincide at $T\rightarrow0$ because the values
$\varepsilon_{Ts0}\simeq42.0\,\mathrm{meV}$ and $\varepsilon_{TsCs}\simeq68.4\,\mathrm{meV}$
were chosen in order to get the same $n_{sc}$ and for higher temperatures
because $\varepsilon_{0}\left(T\right)$ is nearly constant over a
wide temperature range (Fig.~\ref{fig:Influence-of-E0}c)). For the
3D trap density we assumed again $N_{T}=10^{18}\,\mathrm{cm}^{-3}$.

It has to be emphasized that we still use Eq.~(\ref{eq:V-ins-AM}).
But the 2D density of positive charges in the depletion layer $n_{\mathrm{depl}}$
is a byproduct of the calculation of $\varepsilon_{0}$. So it is
no longer a problem to use Eq.~(\ref{eq:V-ins-we}) $V_{\mathrm{ins}}\propto n_{s}+n_{\mathrm{depl}}$
instead of (\ref{eq:V-ins-AM}) $V_{\mathrm{ins}}\propto n_{s}$.
We will do so henceforward. As a result the slope of the line $E_{Ts}-eV_{\mathrm{ins}}Z/D$
in Fig.~\ref{fig:OS-overview-1} is increased, the maximum of $E_{T}\left(Z\right)$
falls off against $\mu$, the $p_{+}\left(Z\right)$-peak gets smaller
and narrower. But it is still possible to let coincide the critical
curves $\rho\left(n_{s}=n_{sc}\right)$ by increasing the trap density
$N_{T}$ in order to compensate the narrower $p_{+}\left(Z\right)$-peak
and by increasing $\varepsilon_{TsCs}$ in order to get the same critical
value $n_{sc}$. Here it is also important that the 2D charge carrier
density of the depletion layer $n_{\mathrm{depl}}\left(T\right)$
has almost no $n_{s}$ dependence for low temperatures as can be seen
in Fig.~\ref{fig:Influence-of-E0}d). The resistivity $\rho\left(T,n_{s}\right)$,
with $N_{T}=2.98\times10^{18}\,\mathrm{cm}^{-3}$ and $\varepsilon_{TsCs}\simeq95.7\,\mathrm{meV}$,
for which we get the same $n_{sc}$ as before, is shown in Fig.~\ref{fig:Influence-of-E0}b).
The above used depletion density $n_{\mathrm{depl}}$ was calculated
under the assumption of a background doping density of $N_{A}=2\times10^{15}\,\mathrm{cm}^{-3}$,
which is a typical value for high mobility Si-MOS samples.%
\footnote{V.\,M.~Pudalov, private communication.%
}  Here the values for $N_{T}$ and $\varepsilon_{TsCs}$ were chosen
in order to get the requested $n_{sc}$ for the given $N_{A}$. In
reality the value $\varepsilon_{TsCs}$ is determined by the chemical
nature of the defect and thus the critical density may be different
from sample to sample if the background doping is different.

If now the slope of the energy $E_{Ts}-eV_{\mathrm{ins}}Z/D$ is higher
due to the inclusion of $n_{\mathrm{depl}}$, the variable $\max\left(E_{T}\left(Z\right)-\mu\right)$
which is crucial for the resistivity is less sensitive on $V_{\mathrm{ins}}$
and $n_{s}$, therefore the transition is less abrupt. The $n_{s}$
dependence of $n_{\mathrm{depl}}$ does not play a role because it
hardly exists.

\section{Spatial Trap Profile\label{sec:Spatial-Trap-Profile}}

At higher temperatures the large value of $k_{B}T$ leads to charged
trap states in regions where the chemical potential is even somewhat
below the chemical potential (i.\,e.\ the $p_{+}\left(Z\right)$-peak
broadens) and therefore the resistivity is increased to unrealistic
high values, see curves at higher temperature in Fig.~\ref{fig:Influence-of-E0}.
But an appreciable density of traps should exist only within the strained
region of the oxide,\citealp{Sze_1981_Physics_of_semiconductor_devices}
and thus the broadening of the peak beyond the width of this region
leads to an unrealistic description. We can resolve this problem by
introducing a spatial trap density profile $N_{T}(Z)$.

If now the trap density $N_{T}$ is a function of $Z$ it has to remain
inside the integral $\Omega_{j}$ (compare Eq.~\ref{eq:Omega-j-1})
\begin{equation}
\Omega_{j}\equiv\int_{0}^{D}N_{T}^{+}\left(Z\right)Z^{j}dZ=\int_{0}^{D}N_{T}\left(Z\right)p_{+}\left(Z\right)Z^{j}dZ.\end{equation}

\begin{figure}
\includegraphics[width=0.8\columnwidth]{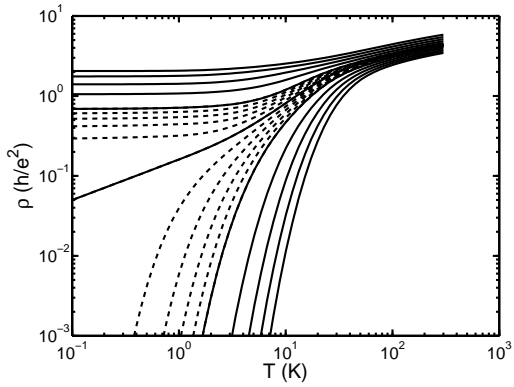}

\caption{\label{fig:rho-T-ns-Zmax}Resistivity $\rho\left(T,n_{s}\right)$
for rectangular spatial trap profile with $Z_{\mathrm{max}}=4\:\mathrm{nm}$
with curves of constant $n_{s}$. The critical density is $n_{sc}=10^{11}\,\mathrm{cm}^{-2}$
which means that $\varepsilon_{TsCs}$ has to be $95.7\,\mathrm{meV}$.
Densities for full lines are $n_{s}=0.75,\dots,1.25\times n_{sc}$
in steps of $0.05\times n_{sc}$, and for dashed lines $n_{s}=0.96,\dots,1.04\times n_{sc}$
in steps of $0.01\times n_{sc}$ (always top down), with $N{}_{T}=n_{T}/Z_{\mathrm{max}}=2.98\times10^{18}\,\mathrm{cm}^{-3}$. }

\end{figure}

For simplicity we use here an rectangular spatial trap profile from
the OS interface to an arbitrary depth $Z_{\mathrm{max}}$, \begin{equation}
N_{T}\left(Z\right)=\begin{cases}
\frac{n_{T}}{Z_{\mathrm{max}}} & \mathrm{for}\,0\le Z\le Z_{\mathrm{max}}\\
0 & \mathrm{for}\, Z>Z_{\mathrm{max}},\end{cases}\end{equation}
where $n_{T}$ is the 2D trap density. Fig.~\ref{fig:rho-T-ns-Zmax}
shows $\rho\left(T,n_{s}\right)$ for $Z_{\mathrm{max}}=4\:\mathrm{nm}$,
the conduction band edge $E_{Cs}$ was used as reference energy, $\varepsilon_{TsCs}\simeq95.7\,\mathrm{meV}$
was chosen in order to get $n_{sc}=10^{11}\,\mathrm{cm}^{-2}$. Where
the 3D trap density $N_{T}$ does not vanish its value is assumed
to be $2.98\times10^{18}\,\mathrm{cm}^{-3}$ as before (see Fig.~\ref{fig:Influence-of-E0}b)),
resulting in $n_{T}=N_{T}\cdot Z_{\mathrm{max}}=1.19\times10^{12}\,\mathrm{cm}^{-3}$
of which again only a part is charged.

As can be seen in Fig.~\ref{fig:rho-T-ns-Zmax}, the behavior for
low temperatures has hardly changed, but for high temperatures $\rho\left(T,n_{s}\right)$
now saturates as a broadening of the $p_{+}(Z)$-peak beyond $Z_{\mathrm{max}}$
does not lead to a further increase in the number of charged scattering
centers. This saturation of $\rho\left(T,n_{s}\right)$ is in fairly
good agreement with experiments, where $\rho$ for high $T$ is limited
as well.

\section{Broadening of the trap energy\label{sec:Broadening-Energy}}

As the trap states will not all be identical and in addition the stochastic
position distribution will influence them mutually, their energetic
position has to be broadened.

We describe the broadening $\Delta E_{T}$ with the help of a normalized
distribution function $g\left(\tilde{E}_{Ts},E_{Ts},\Delta E_{T}\right)$
for the trap energy $E_{Ts}$ which characterizes the trap, see Fig.~\ref{fig:OS-overview-1}.
Now $E_{Ts}$ has the meaning of a mean value. (Mean value should
not be understood in a strict mathematical sense, e.\,g.\ for the
Lorentz distribution the mean value does not exist, but in this case
it is obvious to take the energy $E_{Ts}$ where the distribution
reaches its maximum.) Furthermore $\tilde{E}_{Ts}$ is the value for
a particular trap. The probability of $\tilde{E}_{Ts}$ to lie within
the interval $\left[E,E+dE\right]$ is given by $g\left(E,E_{Ts},\Delta E_{T}\right)dE$.
Therefore we replace the probability $p_{+}$ of a trap to be charged
by\begin{equation}
P_{+}\left(Z\right)=\int_{-\infty}^{\infty}\frac{g\left(\tilde{E}_{Ts},E_{Ts},\Delta E_{T}\right)}{\frac{1}{2}\exp\left(-\frac{\tilde{E}_{Ts}-eV_{\mathrm{ins}}\frac{Z}{D}-\varepsilon_{D}\frac{D}{Z}-\mu}{k_{B}T}\right)+1}d\tilde{E}_{Ts}.\end{equation}
The denominator is that of $p_{+}$, only $E_{Ts}$ is replaced by
$\tilde{E}_{Ts}$. By introducing the dimensionless parameters\begin{align}
\alpha & =\frac{\Delta E_{T}}{k_{B}T},\\
\beta & =\frac{E_{Ts}-eV_{\mathrm{ins}}\frac{Z}{D}-\varepsilon_{D}\frac{D}{Z}-\mu}{k_{B}T}\end{align}
and a dimensionless distribution function $h\left(\eta\right)$ defined
by\begin{align}
\eta & =\frac{\tilde{E}_{Ts}-E_{Ts}}{\Delta E_{T}},\\
g\left(\tilde{E}_{Ts},E_{Ts},\Delta E_{T}\right) & =\frac{1}{\Delta E_{T}}h\left(\frac{\tilde{E}_{Ts}-E_{Ts}}{\Delta E_{T}}\right),\\
g\left(\tilde{E}_{Ts},E_{Ts},\Delta E_{T}\right) & =\frac{1}{\Delta E_{T}}h\left(\eta\right),\end{align}
the probability $P_{+}$ can be written as

\begin{equation}
P_{+}\left(\alpha,\beta\right)=\int_{-\infty}^{\infty}\frac{h\left(\eta\right)}{\frac{1}{2}\exp\left(-\alpha\eta-\beta\right)+1}d\eta.\end{equation}
As a rule this integral cannot be calculated analytically. An exception
from this rule is the uniform distribution. If we define the width
of the 'rectangle' as $2\Delta E_{T}$ we get\begin{equation}
h\left(\eta\right)=\begin{cases}
\frac{1}{2} & \mathrm{for}\,-1<\eta<1\\
0 & \mathrm{elsewhere}\end{cases}\end{equation}
and\begin{equation}
P_{+}\left(\alpha,\beta\right)=\frac{1}{2\alpha}\ln\frac{1+2\exp\left(\beta+\alpha\right)}{1+2\exp\left(\beta-\alpha\right)}.\end{equation}
We also use the normal distribution\begin{equation}
h\left(\eta\right)=\frac{1}{\sqrt{2\pi}}\exp\left(-\frac{\eta^{2}}{2}\right)\end{equation}
with the standard deviation as $\Delta E_{T}$ and the Lorentz distribution
(natural line broadening)\begin{equation}
h\left(\eta\right)=\frac{1}{\pi}\frac{1}{\eta^{2}+1}\end{equation}
with the half full width at half maximum (FWHM) as $\Delta E_{T}$.

\begin{figure*}
\includegraphics[width=0.33\textwidth]{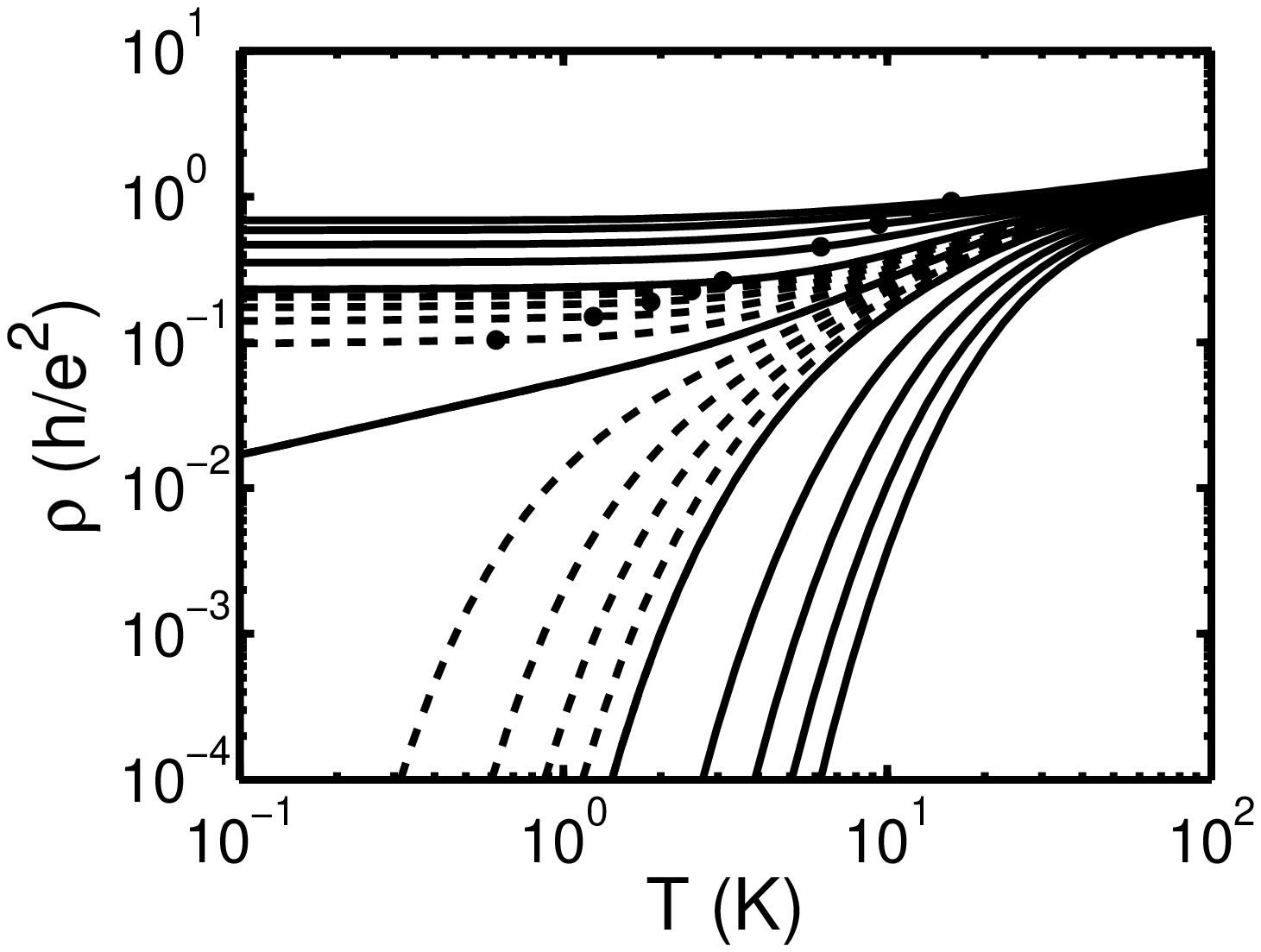}\includegraphics[width=0.33\textwidth]{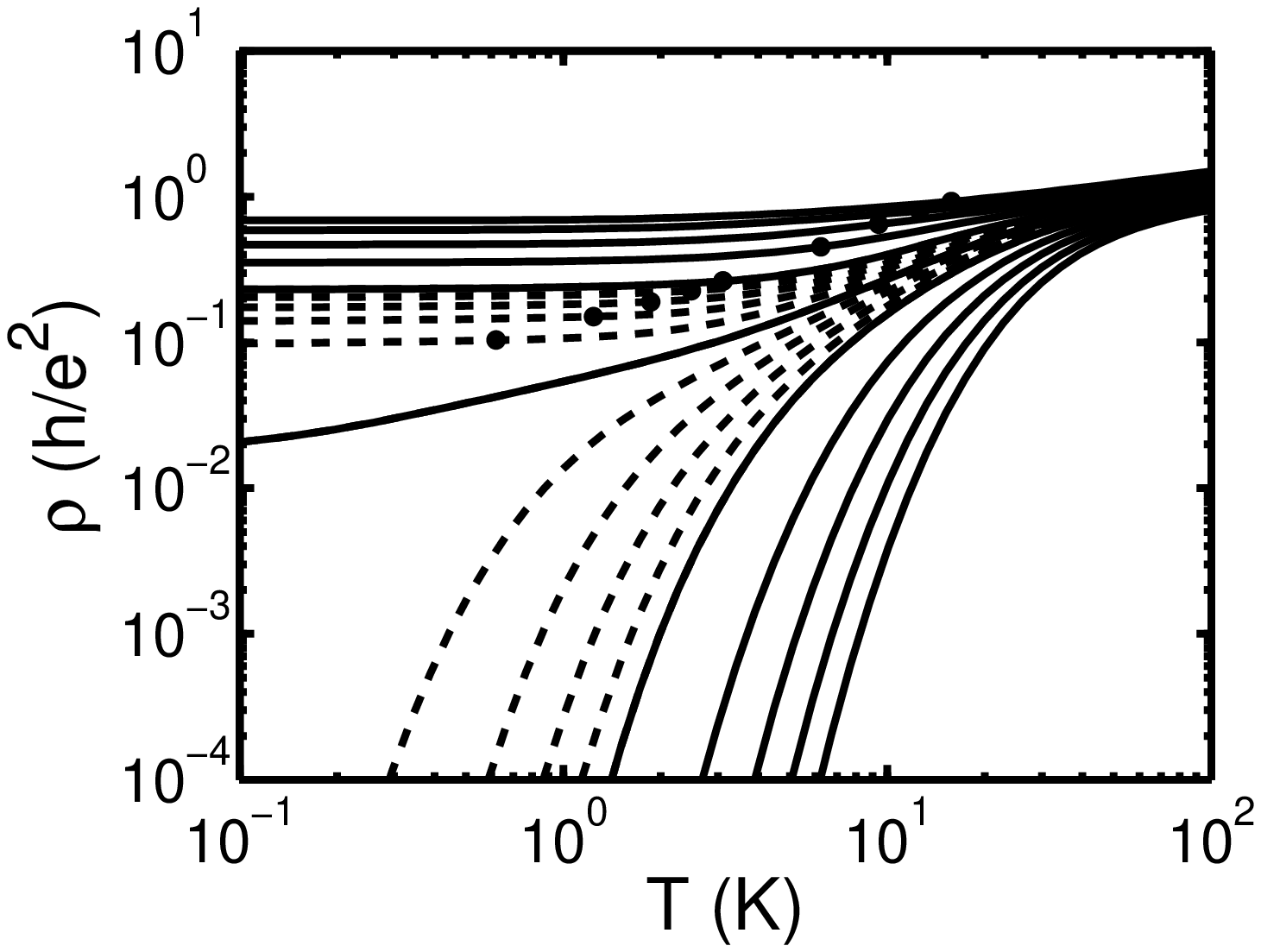}\includegraphics[width=0.33\textwidth]{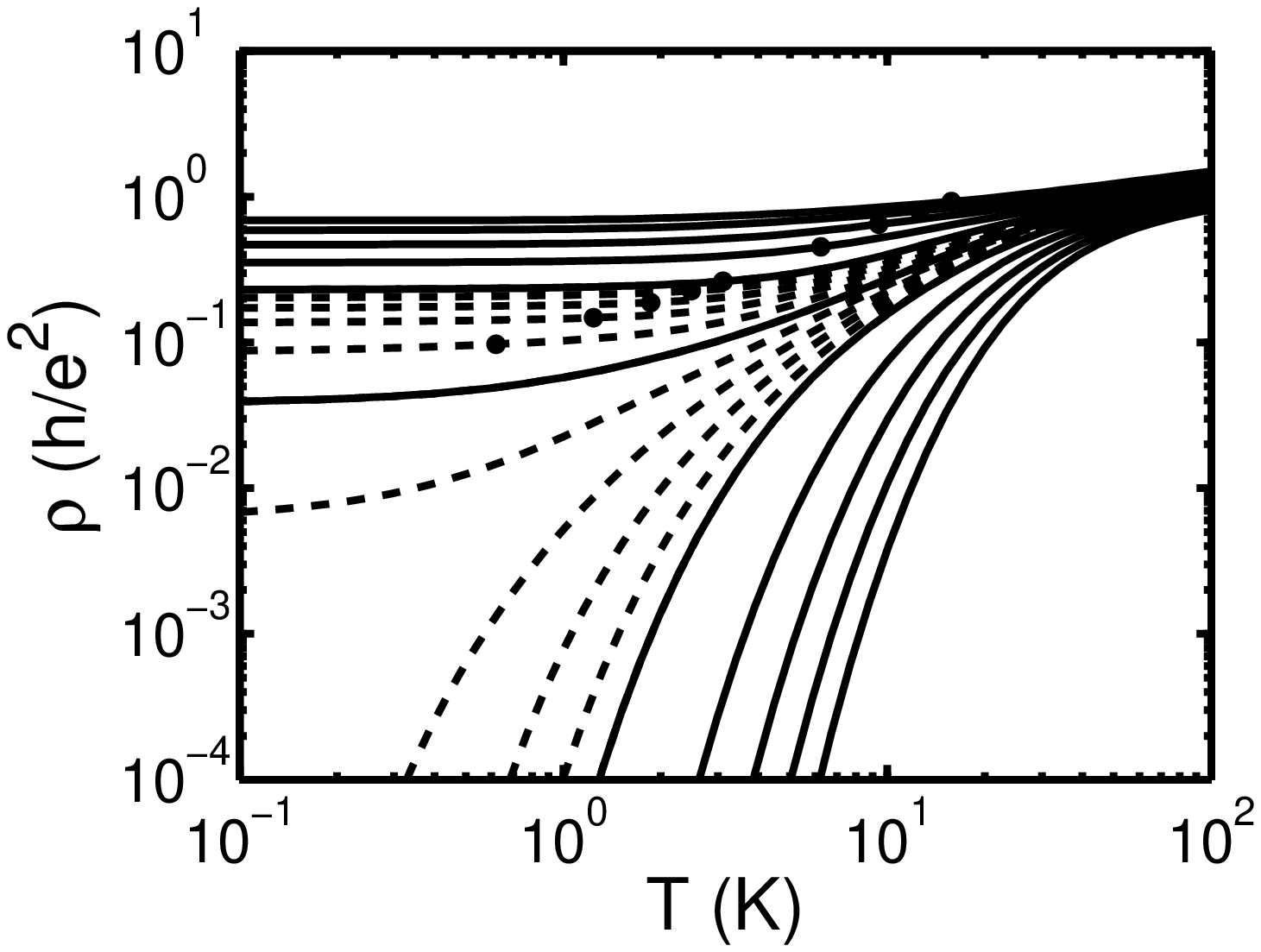}

\includegraphics[width=0.33\textwidth]{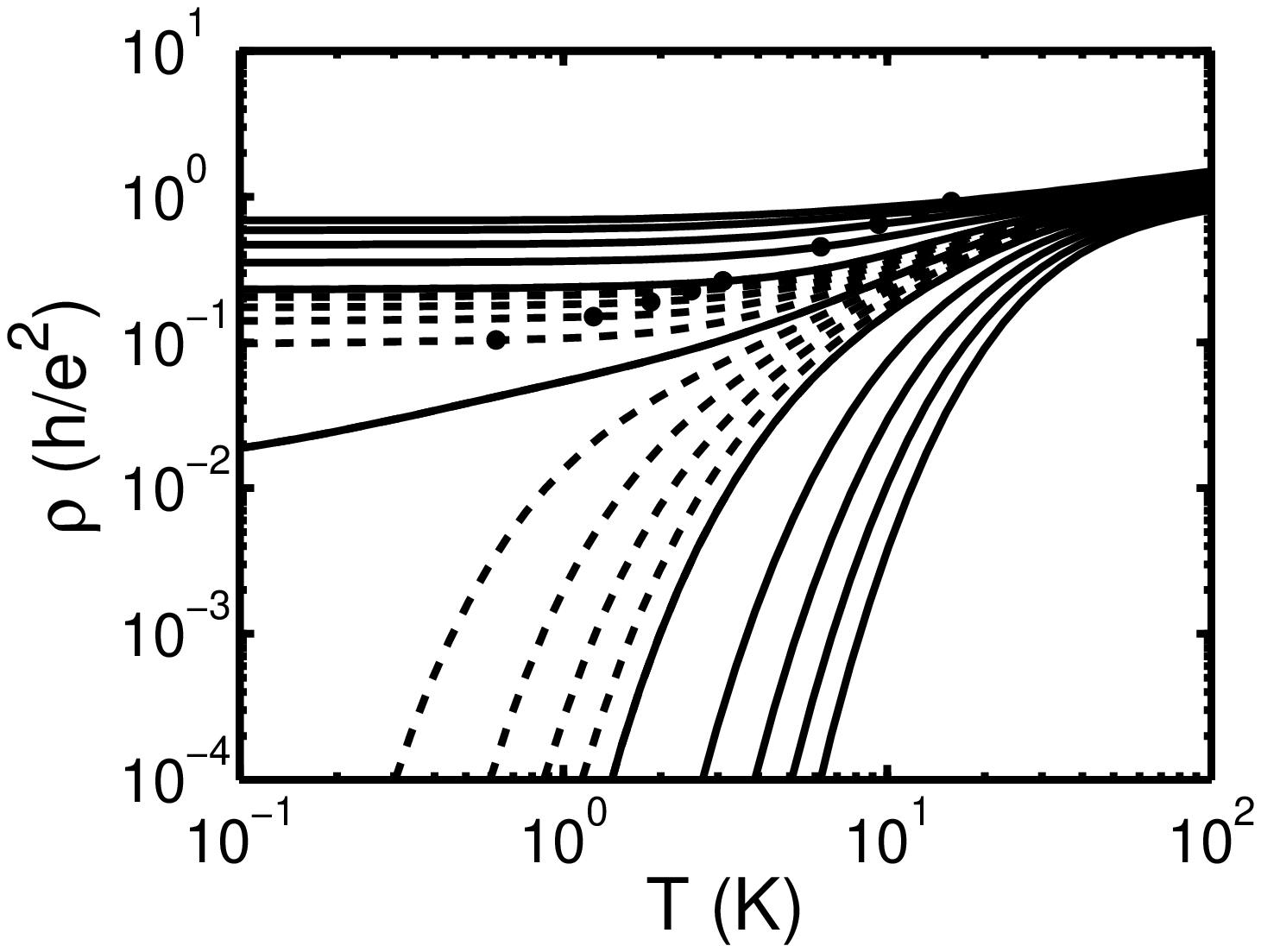}\includegraphics[width=0.33\textwidth]{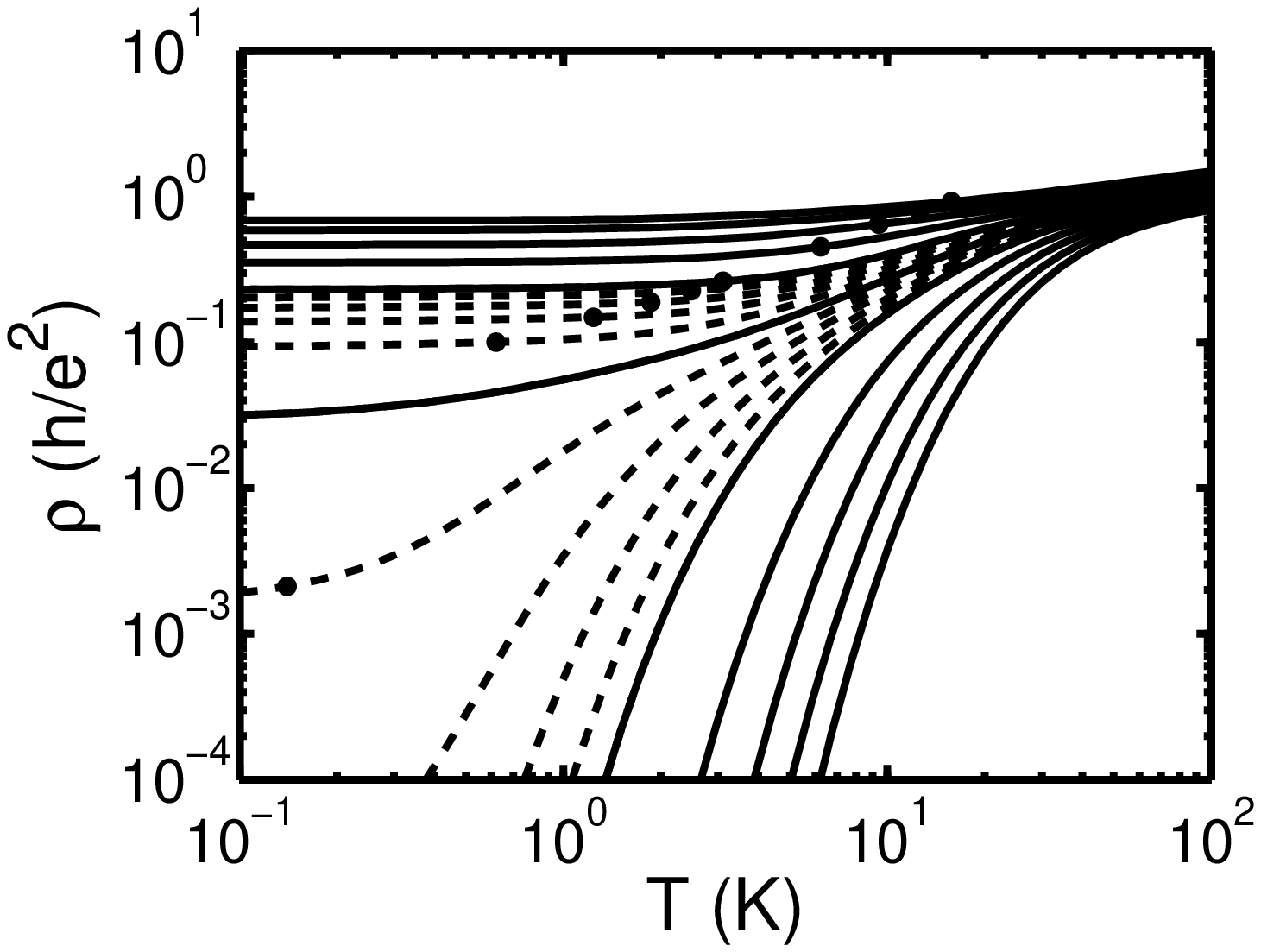}\includegraphics[width=0.33\textwidth]{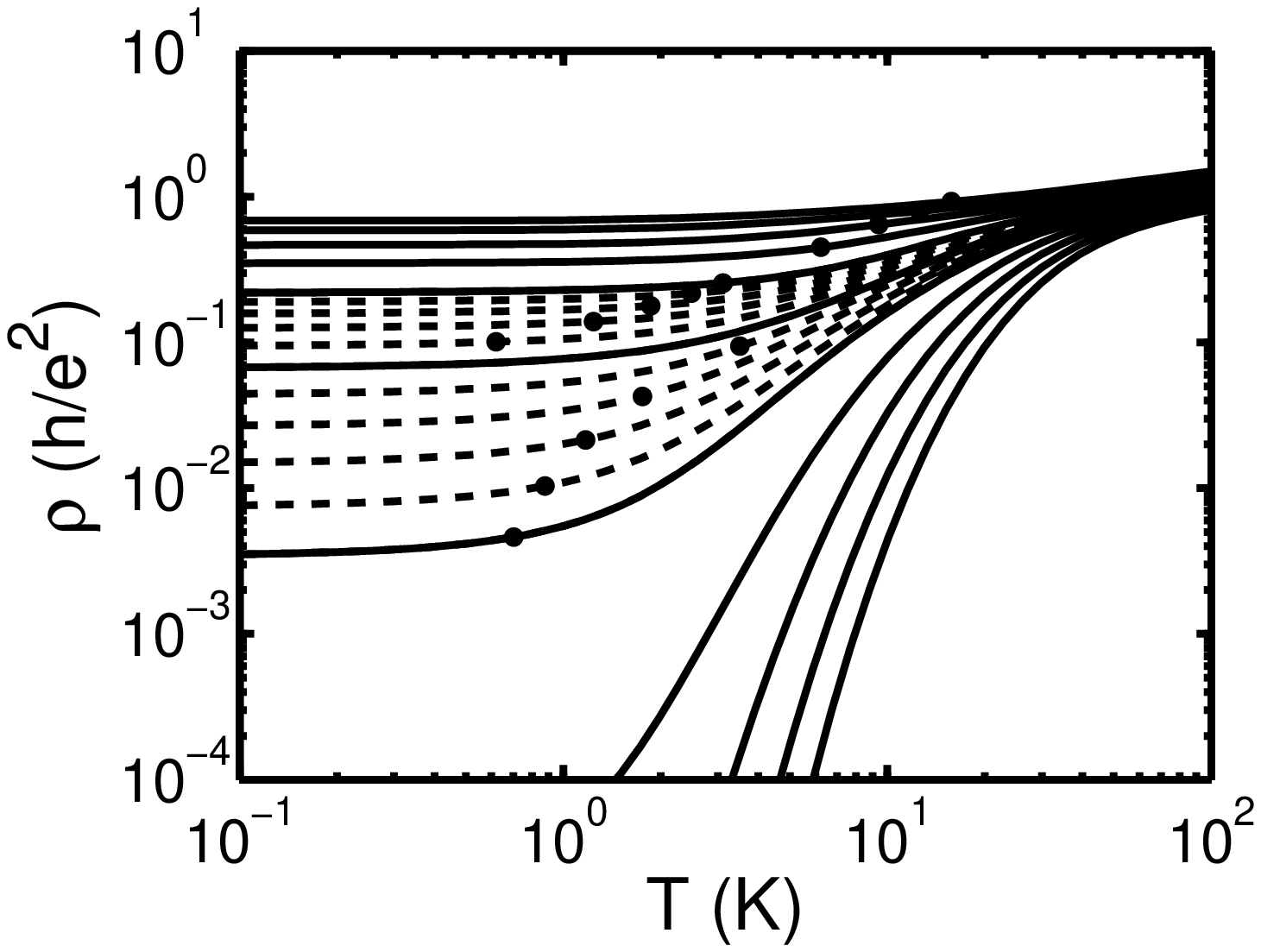}

\includegraphics[width=0.33\textwidth]{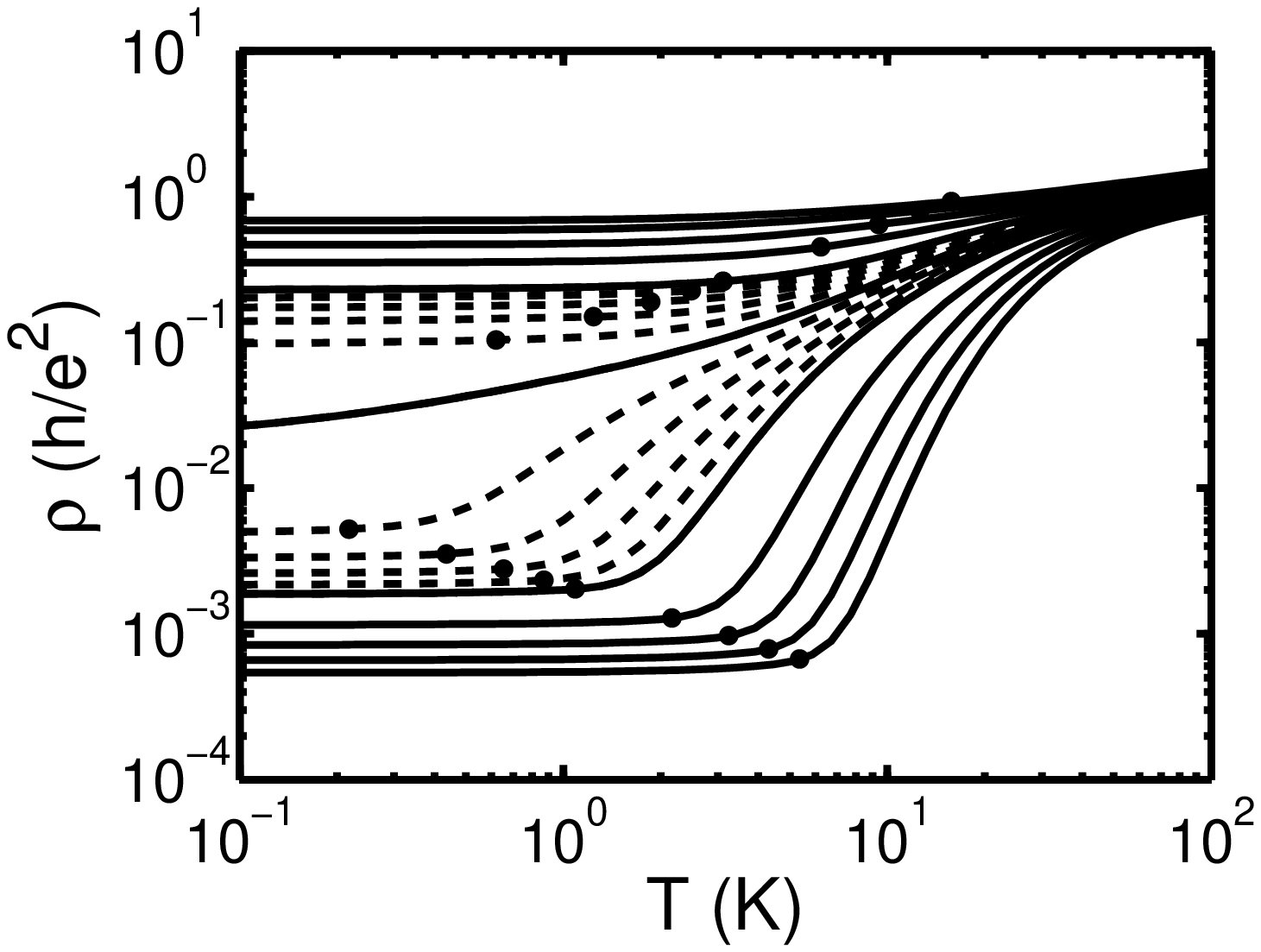}\includegraphics[width=0.33\textwidth]{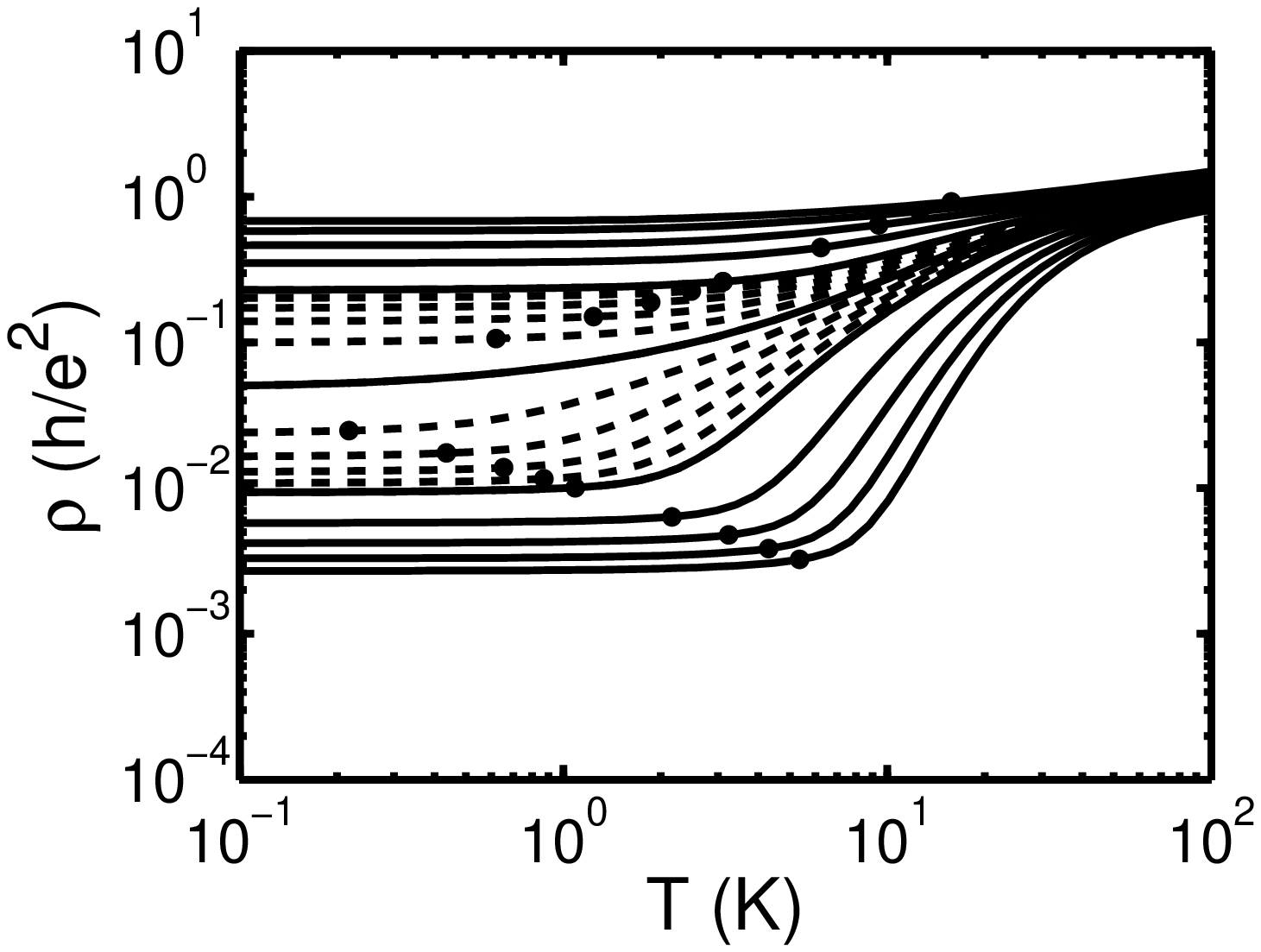}\includegraphics[width=0.33\textwidth]{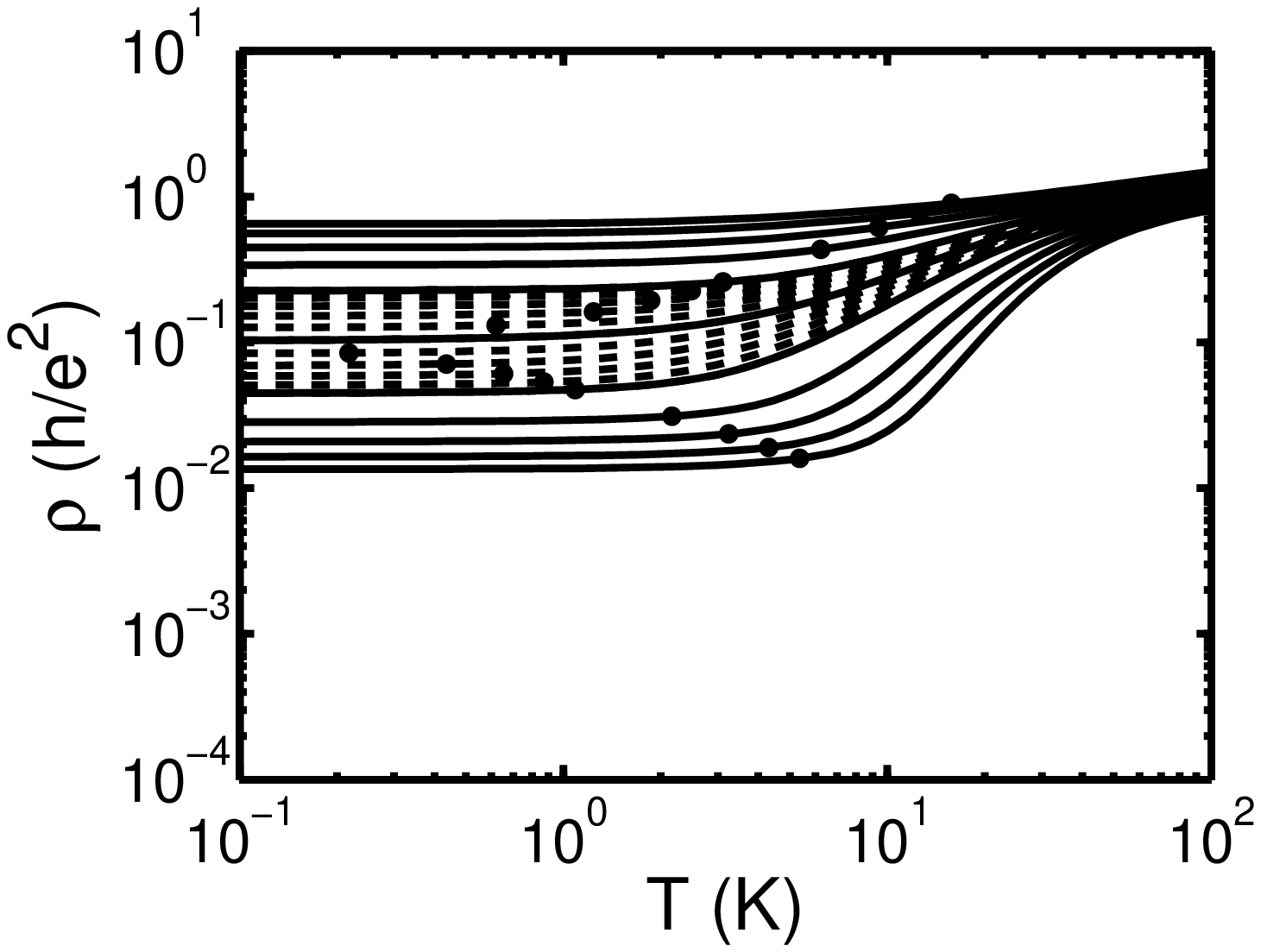}

\caption{\label{fig:broadened-trap-energy}Resistivity $\rho\left(T,n_{s}\right)$
for different types of trap energy broadening functions and different
width $\Delta E_{T}$. The picture columns show uniform distribution,
normal distribution, Lorentz distribution from top to bottom. The
rows show energy width of $\Delta E_{T}=0.02,\,0.1,\,0.5\,\mathrm{meV}$
from left to right. For the curves $n_{s}$ is constant. The critical
density is $n_{sc}=10^{11}\,\mathrm{cm}^{-2}$ for $\varepsilon_{TsCs}\simeq95.7\,\mathrm{meV}$.
Full lines show $n_{s}=0.75,\dots,1.25\times n_{sc}$ in steps of
$0.05\times n_{sc}$ , dashed lines: $n_{s}=0.96,\dots,1.04\times n_{sc}$
in steps of $0.01\times n_{sc}$ (from top to bottom), for a 3D trap
density of $N_{T}=n_{T}/Z_{\mathrm{max}}=10^{18}\,\mathrm{cm}^{-3}$.}

\end{figure*}

Fig.~\ref{fig:broadened-trap-energy} shows the numerical results
for the three different broadening distributions and different width
$\Delta E_{T}=0.02$, $0.1$ and $0.5\,\mathrm{meV}$. Again we took
the conduction band edge at the interface $E_{Cs}$ as reference energy,
assumed that traps exist in the oxide only within $\mathrm{4\, nm}$
from the OS interface (constant trap density within this region),
and used Eq.~(\ref{eq:V-ins-we}) instead of (\ref{eq:V-ins-AM}).
We chose $\varepsilon_{TsCs}\simeq95.7\,\mathrm{meV}$ in order to
get $n_{sc}=10^{11}\,\mathrm{cm}^{-2}$ and for the 3D trap density
$N_{T}=n_{T}/Z_{\mathrm{max}}=10^{18}\,\mathrm{cm}^{-3}$.

As one would expect, the transition is less abrupt for higher $\Delta E_{T}$
and does not vanish even for $\Delta E_{T}=0.5\,\mathrm{meV}$. If
the broadening is indeed caused by potential fluctuations due to disorder
or trap-trap interaction, the value of $\Delta E_{T}$ could even
be much larger than $k_{B}T$ and the transition might be smeared
out even stronger.

In the metallic regime the mean trap energy is below the chemical
potential. As the normal and the Lorentz distribution have tails,
there always remain some charged traps from the upper tail when otherwise
all traps would be filled with electrons and therefore would be neutral.
On a logarithmic resistivity scale the $\rho\left(T\right)$-behavior
is changed drastically by the few additional charged traps.

In the insulating regime the mean trap energy is above the chemical
potential. Here a large part of the traps is charged and by contrast
the few uncharged traps due to the lower tail of the energy distribution
hardly play any role.

By means of analytical considerations  we got the following estimations
for the temperature $T_{b}$ below which the resistivity becomes almost
constant. In general the resistivity $\rho$ varies by some orders
of magnitude, so as a criterion for being almost constant we took
that region where $\rho$ changes finally by only a factor two down
to zero temperature. The markers in Fig.~\ref{fig:broadened-trap-energy}
represent these temperatures $T_{b}$.

According to this definition, for the insulating behavior $n_{s}<n_{sc}$
for all three distributions we get

\begin{equation}
T_{b}=\frac{\max E_{T}\left(Z\right)-E_{F}}{\left(4-\ln2\right)k_{B}}\end{equation}
and for the metallic behavior $n_{s}>n_{sc}$

\begin{equation}
T_{b}=\begin{cases}
\frac{\Delta E_{T}^{2}}{\left(4+\ln2\right)k_{B}\left(E_{F}-\max E_{T}\left(Z\right)\right)} & \dots\mathrm{normal\, distr.}\\
\frac{E_{F}-\max E_{T}\left(Z\right)}{2\left(4+\ln2\right)k_{B}} & \dots\mathrm{Lorentz\, distr.}\end{cases}\end{equation}
The uniform distribution has no tails so in the metallic regime there
is no temperature range where $\rho$ is almost constant.

\section{Conclusions\label{sec:Conclusions}}

In this work we have performed numerical calculations within the dipole
trap model for Si-MOS structures. Originally this model was proposed
by Altshuler and Maslov with several approximations, in order to be
able to get analytical solutions. Due to our numerical treatment we
could eliminate several approximations. We describe the potential
inside the insulator by its detailed spatial dependence instead of
the saddle-point approximation with the quadratic dependence around
its maximum, we fix the trap state energy relative to the conduction
band edge instead of relative to the electronic ground state inside
the triangular potential well and we have taken into account the detailed
change of the chemical potential in the two-dimensional electron layer
with respect to the bulk material, which seems to be more realistic
than the two cases in the original treatment.

According to our calculations, the metallic regime at high electron
densities $n_{s}$, where the resistivity is decreasing towards lower
temperature, is strongly developed . Also a critical density $n_{sc}$
can be identified with a characteristic temperature dependence different
from the metallic and the 'insulating' region. For electron densities
$n_{s}<n_{sc}$, the resistivity curves satturate towards lower temperature
and remain constant when the temperature approaches zero.  They do
not show an insulting behavior in the sense that $\rho$ increases
towards zero temperature. Such an increase can in principle be caused
e.\,g.\ by a further decrease of the chemical potential $\mu$ with
temperature, as in the work of Altshuler and Maslov for case A, where
it was assumed that the temperature dependence of $\mu$ ist the same
for the two-dimensional electron layer as it is in the Si-bulk material.
In the dipol trap model a constant efficient screening is assumed
in order to clarify the effects which are caused when the traps change
their charge state. A realistic treatment of the temperatue dependence
of the electronic screening could also cause an increase in $\rho$
towards lower temperature in that it favors the formation of dipol
trap states. Furthermore, the quantum corrections in the weak and
strong localization regime are also neglected here, but would finally
increase the resistance $\rho$ at low $T$.

In addition, we have further generalized the dipole trap model by
dropping the assumptions that the trap states are homogeneously distributed
inside the oxide layer and that the energy distribution is $\delta$-like.
A narrow spatial distribution of the trap states near the oxide-semiconductor
interface limits the number of charged states at high temperatures
and thus gives an upper limit for the increase of the resistivity
$\rho$ as well. This leads to a good agreement with experimental
observations at higher temperatures.  The energetic broadening of
the trap states on the other hand leads to a finite amount of unoccupied
and thus charged states in cases where otherwise all states would
lie below the chemical potential $\mu$ and the number of charged
trap states would go to zero for $k_{B}T\rightarrow0$. Thus for high
electron densities with metallic behavior the resistivity will not
further decrease towards lower temperature, but saturate at a finite
values, as has been observed in experiments on Si-MOS structures as
well.

The effect of a magnetic field can be taken into account by the Zeeman
splitting of the trap states with spin $\pm1/2$. As shown by Althuler
and Maslov, the energetic splitting can turn a metallic behavior into
an insulating one. We did not include magnetic field effect in our
calculations, but an according energetic shift of the trap states
has to lead to the same effects in our refined model as well.

We also like to mention that for low electron densities care has to
be taken for the dipol scattering model. It is assumed that the electrons
in the two-dimensional layer shield the potential of the charged trap
states and thus form together a dipol field which is responsible for
the scattering. At very low electron densities this screening becomes
weaker and the scattering will finally increase so that the resistivity
should be higher in this regime. These effects have not been taken
into account in the frame of the current work, as we like to present
the basic effects due to charging of trap states.

Althogether, our detailed numerical calculations within the dipol
trap model show that a pronounced metallic state can be caused by
trap states at an appropriate energy level inside the oxide of Si-MOS
structures. For the realistic assumptions of energetic broadening
and narrow spatial distribution near the oxide-semiconductor interface,
the behavior is in close agreement with experimental observations.
\begin{acknowledgments}
Work was supported by the Austrian Science Foundation (FWF) project
no.~P16160. We thank our former colleague A.~Prinz to perform first
calculations within the Altshuler-Maslov Trap model.

\end{acknowledgments}
\appendix

\section{\label{sec:behavior-of-the-chemical-potential}Temperature behavior
of the chemical potential and of the effective electron energy for
low temperatures}

First we show that \begin{equation}
\left.\frac{d^{n}\mu_{E0}\left(T\right)}{dT^{n}}\right|_{\varepsilon_{F0}=\mbox{const.},\, T\rightarrow0}=0\mbox{\quad for }n\ge1\label{eq:mu_E0_T0_2}\end{equation}
holds for $\mu_{E0}\left(T\right)$ from Eq.~(\ref{eq:mu-E0-1}).
In our calculations we use the temperature $T$ and the electron density
in the inversion layer $n_{s}$ as independent variables. So it is
allowed to set $n_{s}=\mbox{const.}$ and therefore also $\varepsilon_{F0}=\mbox{const.}$
while varying T, see Eq.~(\ref{eq:eps-F0}).

For simplicity we introduce the auxiliary variable \begin{equation}
x=\frac{k_{B}T}{\varepsilon_{F0}}\end{equation}
 and the function \begin{equation}
M\left(x\right)=\frac{\mu_{E0}\left(T\left(x\right)\right)-\varepsilon_{F0}}{\varepsilon_{F0}}.\end{equation}
With the chain rule we find\begin{equation}
\frac{d^{n}M\left(x\right)}{dx^{n}}=\frac{\varepsilon_{F0}^{n-1}}{k_{B}^{n}}\frac{d^{n}\mu_{E_{0}}}{dT^{n}}\quad\mathrm{for\,}n\ge1.\end{equation}
So Eq.~(\ref{eq:mu_E0_T0_2}) is equivalent to\begin{equation}
\left.\frac{d^{n}M\left(x\right)}{dx^{n}}\right|_{x\rightarrow0}=0\quad\mathrm{for\,}n\ge1.\label{eq:dnf-dxn-1}\end{equation}
From Eq.~(\ref{eq:mu-E0-1}) we get\begin{equation}
M\left(x\right)=x\ln\left[1-\exp\left(-\frac{1}{x}\right)\right].\end{equation}
The first two derivatives are\begin{align}
\frac{dM}{dx} & =\ln\left[1-\exp\left(-\frac{1}{x}\right)\right]-\frac{1}{\left[\exp\left(\frac{1}{x}\right)-1\right]x},\\
\frac{d^{2}M}{dx^{2}} & =-\frac{1}{\left[\exp\left(\frac{1}{x}\right)-1\right]x^{3}}-\frac{1}{\left[\exp\left(\frac{1}{x}\right)-1\right]^{2}x^{3}}.\end{align}
The second derivative contains only terms of the form\begin{equation}
g_{a,b}=\frac{1}{\left[\exp\left(\frac{1}{x}\right)-1\right]^{a}x^{b}}.\end{equation}
Differentiating $g_{a,b}$ yields\begin{equation}
\frac{dg_{a,b}}{dx}=ag_{a,b+2}+ag_{a+1,b+2}-bg_{a,b+1},\end{equation}
therefore all higher derivatives $\frac{d^{n}M}{dx^{n}}$ also contain
only terms $g_{j,k}$ with $j\in\left\{ 1,2,\dots,n\right\} $ and
$k\in\left\{ n+1,n+2,\dots,2n-1\right\} $. Now we can multiply the
numerator and the denominator of $g_{j,k}$ with $\exp\left(-j/x\right)$
and write\begin{align*}
\lim_{x\rightarrow0}g_{j,k}\left(x\right) & =\lim_{x\rightarrow0}\frac{\exp\left(-\frac{j}{x}\right)}{\left[1-\exp\left(-\frac{1}{x}\right)\right]^{j}x^{k}}\\
 & =\lim_{x\rightarrow0}\frac{\exp\left(-\frac{j}{x}\right)}{x^{k}}\\
 & =\lim_{y\rightarrow\infty}\frac{y^{k}}{j^{k}\exp\left(y\right)},\end{align*}
with $y=j/x$. Applying the rule of L'Hospital $k$ times we get\begin{equation}
\lim_{x\rightarrow0}g_{j,k}\left(x\right)=\lim_{y\rightarrow\infty}\frac{k!}{j^{k}\exp\left(y\right)}=0.\end{equation}
The limit for the first derivative also vanishes\begin{equation}
\lim_{x\rightarrow0}\frac{dM}{dx}=\ln1-\lim_{x\rightarrow0}g_{1,1}=0,\end{equation}
so Eq.~(\ref{eq:dnf-dxn-1}) holds and \begin{equation}
\mu_{E_{0}}\simeq E_{F0}\quad\mathrm{for}\, k_{B}T\ll\varepsilon_{F0}.\end{equation}

With help of Eq.~(\ref{eq:mu_E0_T0_2}) we show now that\begin{equation}
\left.\frac{d^{n}\bar{\varepsilon}\left(T\right)}{dT^{n}}\right|_{\varepsilon_{F0}=\mbox{const.},\, T\rightarrow0}=0\mbox{\quad for }n\ge1\end{equation}
holds for Eq.~(\ref{eq:epsilon-bar-6}).

For not too small $\eta$ the Fermi-Dirac integral $\mathcal{F}_{j}\left(\eta\right)$
can be approximated by\citealp{Blakemore_Semiconductor_Statistics}\begin{equation}
\mathcal{F}_{j}\left(\eta\right)\simeq\frac{\eta^{j+1}}{\Gamma\left(j+2\right)}.\end{equation}
For low temperatures $\mu_{E_{0}}$ approaches $\varepsilon_{F0}$
and therefore it is positive, so from Eq.~(\ref{eq:epsilon-bar-6})
we get\begin{equation}
\bar{\varepsilon}\simeq\varepsilon_{F0}\left(\frac{\varepsilon_{F0}}{\mu_{E_{0}}}\right)^{5}\quad\mathrm{for}\, k_{B}T\ll\varepsilon_{F0},\label{eq:epsilon-bar-7}\end{equation}
and with the chain rule respectively Faá di Bruno's formula\begin{align}
\frac{d\bar{\varepsilon}}{dT} & =\frac{d\bar{\varepsilon}}{d\mu_{E_{0}}}\frac{d\mu_{E_{0}}}{dT},\\
\frac{d^{n}\bar{\varepsilon}}{dT^{n}} & =\sum\frac{n!}{k_{1}!\dots k_{n}!}\frac{d^{k}\bar{\varepsilon}}{d\mu_{E_{0}}^{k}}\nonumber \\
 & \quad\quad\times\left(\frac{1}{1!}\frac{d\mu_{E_{0}}}{dT}\right)^{k_{1}}\dots\left(\frac{1}{n!}\frac{d^{n}\mu_{E_{0}}}{dT^{n}}\right)^{k_{n}},\label{eq:dnepsbar-dTn}\\
k & =k_{1}+\dots+k_{n},\nonumber \end{align}
where the sum runs over all integer numbers $k_{1},\dots,k_{n}\ge0$
which fulfill\begin{equation}
k_{1}+2k_{2}+\dots+nk_{n}=n.\end{equation}
We do not have to find this numbers, we only need to know that for
$n>1$ at least one $k_{j}>0$, therefore each term in (\ref{eq:dnepsbar-dTn})
contains a factor $\frac{d^{j}\mu_{E_{0}}}{dT^{j}}$, so for $n_{s}=\mathrm{const.}$,
$\mathrm{T\rightarrow0}$ the sum vanishes and from Eq.~(\ref{eq:epsilon-bar-7})
we get\begin{equation}
\bar{\varepsilon}\simeq\varepsilon_{F0}\quad\mathrm{for}\, k_{B}T\ll\varepsilon_{F0}.\end{equation}

\section{\label{sec:Ground-state-energy}Ground state energy of the inversion
layer}

As mentioned before we calculate the ground state energy of the inversion
layer with help of the Ritz variational principle using the Fang-Howard
test envelope wave function\citealp{Fang_Howard_1966}\begin{equation}
\varphi\left(z,b\right)=\begin{cases}
\sqrt{\frac{b^{3}}{2}}z\exp\left(-\frac{bz}{2}\right) & \mathrm{for}\, z\ge0\\
0 & \mathrm{for}\, z<0\end{cases}\end{equation}
and as an approximation for the potential\begin{align}
U\left(z\right) & =\begin{cases}
U_{d}\left(z\right)+U_{s}\left(z\right)+U_{i}\left(z\right) & \mathrm{for}\, z\ge0\\
\infty & \mathrm{for}\, z<0,\end{cases}\\
U_{d}\left(z\right) & \simeq\begin{cases}
\frac{e^{2}n_{\mathrm{depl}}}{\epsilon_{\mathrm{sc}}\epsilon_{0}}z\left(1-\frac{z}{2z_{d}}\right) & \mathrm{for}\, z<z_{d}\\
\frac{e^{2}n_{\mathrm{depl}}}{2\epsilon_{\mathrm{sc}}\epsilon_{0}}z_{d} & \mathrm{for}\, z>z_{d},\end{cases}\label{eq:Ud-1}\\
U_{s}\left(z\right) & \simeq\frac{e^{2}n_{s}}{2b\epsilon_{\mathrm{sc}}\epsilon_{0}}\left[6-\left(\left(bz\right)^{2}+4bz+6\right)\exp\left(-bz\right)\right],\\
U_{i}\left(z\right) & \simeq\frac{be^{2}}{32\pi\epsilon_{\mathrm{sc}}\epsilon_{0}}\frac{\epsilon_{\mathrm{sc}}-\epsilon_{\mathrm{ins}}}{\epsilon_{\mathrm{sc}}+\epsilon_{\mathrm{ins}}}\frac{1}{4z}.\end{align}
The term $U_{d}$ comes from the charged acceptors within the depletion
layer with thickness $z_{d}$, $U_{s}$ describes the interaction
with all other electrons in the inversion layer, and $U_{i}$ the
interaction with image charges. To write $U_{s}\left(z\right)$ in
this form we have to assume that only the first subband is occupied,
this is the so called quantum limit. The conduction band edge is built
by $U_{d}\left(z\right)+U_{s}\left(z\right)$, therefore the zero
point of the energy scale was chosen to get $U_{d}\left(0\right)+U_{s}\left(0\right)=0$
which means that the resulting ground state energy is measured against
the conduction band edge at the interface $E_{Cs}$ as requested.

The Hamiltonian is given by $\hat{H}=\hat{T}+U$ with the operator
for the kinetic energy $\hat{T}=-\frac{\hbar^{2}}{2m_{z}}\frac{\partial^{2}}{\partial z^{2}}$,
where $m_{z}$ is the $z$-component of the effective mass of the
semiconductor in the bulk. The ground state energy is the expectation
value of the Hamiltonian,\begin{equation}
\varepsilon_{0}=\Braket{\hat{T}}+\Braket{U_{d}\vphantom{\hat{T}}}+\Braket{U_{s}\vphantom{\hat{T}}}+\Braket{U_{i}\vphantom{\hat{T}}},\end{equation}
calculated with the value of $b$ which makes the total energy per
electron \begin{equation}
\tilde{\varepsilon}=\Braket{\hat{T}}+\Braket{U_{d}\vphantom{\hat{T}}}+\frac{1}{2}\Braket{U_{s}\vphantom{\hat{T}}}+\Braket{U_{i}\vphantom{\hat{T}}}\end{equation}
minimal. The factor $1/2$ in the third term prevents from double
counting the electron-electron interaction. With the density $n^{*}=n_{\mathrm{depl}}+\frac{11}{32}n_{s}$
introduced by AFS\citealp{AFS_RevModPhys.54.437} one gets\begin{align}
\tilde{\varepsilon} & =\frac{\alpha}{2}b^{2}+\beta b+\gamma b^{-1}-\frac{\delta\left(b\right)}{2}b^{-2},\\
\alpha & =\frac{\hbar^{2}}{4m_{z}},\\
\beta & =\frac{e^{2}}{32\pi\epsilon_{\mathrm{sc}}\epsilon_{0}}\frac{\epsilon_{\mathrm{sc}}-\epsilon_{\mathrm{ins}}}{\epsilon_{\mathrm{sc}}+\epsilon_{\mathrm{ins}}},\\
\gamma & =\frac{3e^{2}n^{*}}{\epsilon_{\mathrm{sc}}\epsilon_{0}},\\
\delta & =\frac{12e^{2}\left(N_{A}-N_{D}\right)}{\epsilon_{\mathrm{sc}}\epsilon_{0}}\\
 & \quad\quad\times\left\{ 1-\left[\frac{\left(bz_{d}\right)^{2}}{12}+\frac{bz_{d}}{2}+1\right]\exp\left(-bz_{d}\right)\right\} .\end{align}
$N_{A}$ and $N_{D}$ are the densities of the acceptors and donors
respectively. The coefficients have been chosen in order to get $\alpha,\,\beta,\,\gamma,\,\delta>0$
and to get a most simple equation\begin{equation}
\frac{d\tilde{\varepsilon}}{db}=\alpha b+\beta-\gamma b^{-2}+\delta\left(b\right)b^{-3}=0.\end{equation}
As AFS used $\frac{e^{2}n_{\mathrm{depl}}}{\epsilon_{\mathrm{sc}}\epsilon_{0}}z\left(1-\frac{z}{2z_{d}}\right)$
for $0<z<\infty$ instead of Eq.~(\ref{eq:Ud-1}) they did not get
the term proportional to $\exp\left(-bz_{d}\right)$, under normal
circumstances it is very small but formally it is necessary to see
that $b\rightarrow0$, $\tilde{\varepsilon}\rightarrow-\infty$ is
not the global minimum. Furthermore they neglected $\beta$ and $\delta\left(b\right)b^{-3}$
and got\begin{equation}
b=\left(\frac{\gamma}{\alpha}\right)^{1/3}=\left(\frac{12e^{2}n^{*}m_{z}}{\epsilon_{\mathrm{sc}}\epsilon_{0}\hbar^{2}}\right)^{1/3}.\end{equation}
We neglected only $\delta\left(b\right)b^{-3}$, this leads to\begin{align}
b & =F\left(K+\frac{1}{K}-1\right),\\
F & =\frac{\beta}{3\alpha}=\frac{e^{2}m_{z}}{24\pi\epsilon_{\mathrm{sc}}\epsilon_{0}\hbar^{2}},\\
K & =\left[\frac{1}{2}\left(3B-2+\sqrt{3B\left(3B-4\right)}\right)\right]^{1/3},\\
B & =\frac{9\alpha^{2}\gamma}{\beta^{3}}=\frac{\gamma}{\beta F^{2}}=\frac{96\pi n^{*}}{F^{2}}\frac{\epsilon_{\mathrm{sc}}+\epsilon_{\mathrm{ins}}}{\epsilon_{\mathrm{sc}}-\epsilon_{\mathrm{ins}}}.\end{align}

The density $n_{\mathrm{depl}}$ can be calculated from the total
band bending $e\phi_{0}=E_{Cb}-E_{Cs}$ (b=bulk, s=surface),\begin{align}
en_{\mathrm{depl}} & =e\left(N_{A}-N_{D}\right)z_{d}=\sqrt{2\left(N_{A}-N_{D}\right)\epsilon_{\mathrm{sc}}\epsilon_{0}e\phi_{0\mathrm{d}}},\\
e\phi_{0d} & =e\phi_{0}-e\phi_{0s}-k_{B}T.\end{align}
Here $e\phi_{0\mathrm{d}}$ is the band bending caused by the charges
in the depletion layer (acceptors and donors) and $e\phi_{0s}$ that
caused by the free electrons in the inversion layer. To get this equations
one has to assume that in the depletion layer all acceptors are charged
and that in the bulk there is charge neutrality. The boundary between
the depletion layer and the bulk is not sharp, this is described by
the term $-k_{B}T$ in $e\phi_{0d}$.\citealp{Stern_1972_Self-Consistent_Results}
To calculate $\phi_{0s}$ one has to solve the Poisson equation with
the charge density which corresponds with the test wave function $\varphi$,
the result is\begin{equation}
\phi_{0s}=\frac{3en_{s}}{b\epsilon_{\mathrm{sc}}\epsilon_{0}}.\end{equation}

As can be seen from Fig.~\ref{fig:OS-overview-2} for the total band
bending\begin{align}
e\phi_{0} & =E_{Cb}-E_{Cs},\nonumber \\
e\phi_{0} & =E_{Cb}-\mu+\mu-E_{0}+E_{0}-E_{Cs},\nonumber \\
e\phi_{0} & =-\mu_{Cb}+\mu_{E_{0}}+\varepsilon_{0}\end{align}
holds. The chemical potential relative to the conduction band edge
in the bulk $\mu_{Cb}$ is determined by the charge neutrality and
can be calculated by solving\begin{multline}
N_{C}\mathcal{F}_{1/2}\left(\frac{\mu_{Cb}}{k_{B}T}\right)+\frac{N_{A}}{g_{A}\exp\left(\frac{\varepsilon_{A}-\varepsilon_{g}-\mu_{Cb}}{k_{B}T}\right)+1}=\\
=N_{V}\mathcal{F}_{1/2}\left(\frac{-\varepsilon_{g}-\mu_{Cb}}{k_{B}T}\right)\end{multline}
numerically. $N_{C}$ and $N_{V}$ are the effective densities of
state in the conduction band and in the valence band, $\varepsilon_{A}$
is the acceptor ionization energy, $g_{A}$ the acceptor degeneracy
factor, and $\varepsilon_{g}$ the gap energy. According to Sze\citealp{Sze_1981_Physics_of_semiconductor_devices}
this quantities are given by\begin{align}
N_{C} & =g_{s}g_{\mathrm{v3D}}\left(\frac{m_{\mathrm{de3D}}k_{B}T}{2\pi\hbar^{2}}\right)^{3/2},\\
N_{V} & =g_{s}\left(\frac{m_{\mathrm{dh3D}}k_{B}T}{2\pi\hbar^{2}}\right)^{3/2},\\
\varepsilon_{g} & =\varepsilon_{\mathrm{g0Si}}-\frac{\alpha_{\mathrm{Si}}T^{2}}{T+\beta_{\mathrm{Si}}}.\end{align}
The values we used can be found in Table \ref{tab:values}.

The ground state energy $\varepsilon_{0}$ itself is a (small) part
of the the total band bending, this problem is solved by a fix point
iteration using $\varepsilon_{0}=0$ as start value.

\bibliographystyle{apsrev4-1}
\bibliography{paper4sub}

\end{document}